\pdfoutput=1
\documentclass[preprint,3p,twocolumn]{elsarticle}
\usepackage{graphicx}
\usepackage{mathrsfs}
\usepackage{algorithm,algorithmicx, algpseudocode}
\usepackage{amsmath,amssymb}
\usepackage{caption}
\usepackage{subcaption}
\usepackage{multicol}
\usepackage[export]{adjustbox}
\usepackage{lineno,hyperref}
\usepackage{fancyvrb}
\usepackage{bchart}
\usepackage{booktabs}
\usepackage{tikz}
\usetikzlibrary{arrows.meta,positioning,shapes.geometric,calc}
\usepackage{pgfplots}
\pgfplotsset{compat=1.18}

\modulolinenumbers[5]

\def\CircleArrow{\hbox{$\circ$}\kern-1.5pt\hbox{$\rightarrow$}}

\algnewcommand\algorithmicnot{\textbf{not}}
\algdef{SE}[IF]{IfNot}{EndIf}[1]{\algorithmicif\ \algorithmicnot\ #1\ \algorithmicthen}{\algorithmicend\ \algorithmicif}%

\journal{arXiv}

\let\today\relax
\makeatletter
\def\ps@pprintTitle{%
    \let\@oddhead\@empty
    \let\@evenhead\@empty
    \def\@oddfoot{\footnotesize\itshape
         {~} \hfill\today}%
    \let\@evenfoot\@oddfoot
    }
\makeatother

\newcommand{\MR}{MarketRank}
\newcommand{\FS}{Fluxscape}

\begin{document}

\begin{frontmatter}

\title{MarketRank: ranking stocks by the stationary probabilities of a Markov chain over estimated money flows, with a topological landscape view}

\author{B. Kaan Karamete\corref{cc}}

\cortext[cc]{\it Corresponding author: B. Kaan Karamete, karametebkaan@gmail.com}
\address{Graph AI LLC}

\begin{abstract}
Every trading day, money leaves some stocks and moves into others. \MR\ asks where that money concentrates and how the concentration shifts over time. It treats the market as a Markov chain whose states are stocks. The chance of moving from one stock to another is the share of the first stock's outgoing dollars that goes to the second, and a stock's MarketRank score is its long-run (stationary) probability in that chain, computed by damped power iteration on a sparse transition matrix. PageRank ranks web pages with the same idea, using links instead of dollars.

Who sold what to buy what is not public, so the flows are \emph{estimated} from daily prices and volumes with a correlation-tilted gravity model, in a deterministic parallel C++ engine. On a universe of 10{,}000 US stocks and ETFs, the ranking is led by the large semiconductor and technology names, and estimated flows stay within a sector more often than chance. The level of the score, however, does not predict next-day returns: it shows where money sits, not where prices go next. A 3D landscape view, \FS, turns the ranking into terrain: Louvain communities of the recent flow graph, re-clustered on every bar, become territories, ordered by recursive spectral bisection along a generalized Hilbert curve, interpolated by inverse-distance weighting, smoothed by a centroidal-Voronoi-style relaxation with portfolio holdings pinned to their exact values, and rendered as a Gouraud-shaded mesh. The surface is for reading trends; the exact ranked table remains the reference.

The practical question is whether MarketRank can improve real portfolio moves, and we answer it with tests fixed in advance. First, a ten-year walk-forward: step through 2016--2026 week by week, decide each week's moves using only past data, and ask whether following MarketRank beats simply holding the current portfolio after trading costs. It does not. Every configuration lost 3--5\% a year against buy-and-hold, and the one persistent effect, a short-term reversal in which today's hottest names do slightly worse, is statistically real but economically small. Second, we compare the estimated flows with \emph{observed} ones: the quarter-to-quarter holding changes of every institutional manager reporting to the SEC on Form 13F, 39 quarters and 4,000--9,000 managers each. The estimate knows roughly \emph{which} stocks send and receive money and a little about \emph{when}, but nothing about \emph{who pairs with whom} beyond those totals, and plain trading volume ranks the observed destinations better than the estimated score does. Third, the flow communities that shape the landscape are re-clustered from scratch on every bar and measured: they persist only about as long as the flow window that defines them. The conclusion is that the Markov solver is sound but its transition probabilities are only as good as the connectivity fed into it: public daily bars carry size and a little timing, not pairing. Three further tests, each registered before the data it judges, ask whether the connectivity can be repaired: learning it from weekly data, learning a correction to the chain from 15-minute bars, and testing the chain as it is on two years of real 15-minute data for the 1,000 most liquid names. All three are negative. Neither a learned connectivity nor the estimated one predicts returns better than graph-free controls, and the chain does no better than the same chain attached to the wrong stocks.
\end{abstract}


\end{frontmatter}

\section{Introduction}
\label{Section:intro}

Consider a fund that trims a large technology position on Monday and spends the proceeds on two semiconductor makers and a bond ETF. Seen from outside, three prices moved and volume rose; seen from inside, a definite amount of money left one stock and arrived at three others. Multiply that by thousands of managers, index rebalances and retail rotations, and the market's daily trading becomes a directed, weighted network: each edge says how many dollars moved \emph{from} one name \emph{to} another. Two questions follow naturally. Where does that money settle when the flows are followed to their long-run end? And how does that settling point rotate from week to week?

A finite-state Markov chain~\cite{norris,kemenysnell} answers exactly this kind of question. If the states are stocks and the probability of moving from $j$ to $i$ is the share of $j$'s outflow that goes to $i$, then the chain's stationary distribution $\pi$ is the long-run fraction of money at each stock. \MR\ is a solver for that stationary distribution. Let $T_{j\to i}\ge 0$ be the dollars that moved from selling stock $j$ into buying stock $i$ over a window. The transition probability from $j$ to $i$ is the out-share $T_{j\to i}/\sum_k T_{j\to k}$. With damping $p$, the stationary condition of the damped chain is the \emph{MarketRank Condition} (MRC)
\begin{equation}
\begin{aligned}
r_i&=(1-p)\sum_{j\in B(i)}P_i(r_j)+\frac{p}{N},\\
P_i(r_j)&=r_j\,\frac{T_{j\to i}}{\sum_k T_{j\to k}},
\end{aligned}
\label{eqn:mrc}
\end{equation}
where $B(i)$ is the set of stocks that send money to $i$, $N$ the number of active stocks and $p=0.15$. Its solution $r=\pi$ is the MarketRank score. PageRank~\cite{brinpage,pagerank} is the best-known instance of the same construction: a damped chain whose transition probabilities are the out-shares of hyperlinks. MarketRank is analogous, with dollars in place of links. A stock ranks high when it receives a large share of what its senders give away, and when those senders themselves rank high. The rank is a property of the whole flow network, not of a stock's own volume.

The difficulty is that trades are anonymous: a tape shows that a stock traded, not which stock the buyer sold to fund it. \MR\ therefore \emph{estimates} $T$ from public bars, and is explicit that every downstream number inherits that estimate.

\paragraph{Contributions}
\begin{itemize}
\item \textbf{The model:} a damped Markov chain over stocks whose transition probabilities are dollar out-shares. Its stationary distribution, defined by the MRC~(\ref{eqn:mrc}), is the score. Dangling stocks are handled by teleport. A worked three-stock example reproduced by the test suite (Section~\ref{Section:model}).
\item \textbf{Flow estimation from public bars:} signed dollar pressure, a correlation-tilted gravity pairing with exact row and column totals in $O(NW)$, sparse candidate sinks, cumulative accumulation and two-sided pruning (Section~\ref{Section:flows}).
\item \textbf{A scalable deterministic engine:} C++20 and OpenMP with bit-identical results across thread counts, a CSR matrix and its transpose, and a DuckDB catalogue over a Hive-partitioned Parquet lake fed from the Alpaca SIP feed, at 10,000 tickers (Section~\ref{Section:engine}).
\item \textbf{The \FS\ landscape:} a deterministic pipeline from the flow graph to a shaded 3D terrain, with a measurement of how long its communities actually persist (Section~\ref{Section:landscape}).
\item \textbf{A pre-registered walk-forward test} over ten years of weekly decisions, after costs and with multiple-testing control (Section~\ref{Section:wf}).
\item \textbf{A validation against observed flows} from SEC Form 13F, with size, temporal and support-matched gravity nulls (Section~\ref{Section:13f}).
\item \textbf{Three pre-registered tests of the connectivity}, weekly and at 15 minutes, with graph-free, identity and relabeled-chain controls and power curves on synthetic markets with a planted graph (Section~\ref{Section:learned}).
\end{itemize}
Section~\ref{Section:results} reports a first evaluation on real data, Section~\ref{Section:wf} the walk-forward test, Section~\ref{Section:13f} the comparison with 13F, Section~\ref{Section:learned} the tests of learned and intraday connectivity, and Section~\ref{Section:limits} the limitations and planned work.

\section{Related work}
\label{Section:related}

\paragraph{Markov chains and stationary distributions} An irreducible, aperiodic finite Markov chain has a unique stationary distribution $\pi=\pi P$, and power iteration converges to it~\cite{norris,kemenysnell}. Damping, that is mixing $P$ with a uniform jump of probability $p$, makes any chain irreducible and aperiodic, and the convergence rate is then at most $1-p$ per iteration. \MR\ is a solver for this stationary distribution, on a chain whose transition probabilities are estimated dollar out-shares.

\paragraph{PageRank as an instance} PageRank~\cite{brinpage,pagerank} is the best-known application of the same transition-probability concept. A random surfer follows out-links with probability $1-p$ and teleports with probability $p$, and dangling pages are usually redistributed uniformly. \MR\ uses the same damped chain with dollar out-shares in place of link out-shares. Like PageRank, MarketRank is a Markov-chain ranking.

\paragraph{The author's Markov-chain lineage} This work continues the author's earlier use of Markov-chain transition probabilities:
\begin{itemize}
\item an adaptive Markov chain for map matching of vehicle GPS trips~\cite{kineticamapmatching};
\item the probability-rank and PageRank graph solvers of Kinetica-Graph~\cite{kineticagraph,kinetica}.
\end{itemize}
The \MR\ solver is native C++ with no database dependency.

\paragraph{Money-flow indicators} Technical money-flow indicators combine price change and volume \emph{per stock}, so a stock's flow is a scalar with no counterparty. \MR\ keeps the same raw ingredients (return, volume, VWAP) but pairs sources with sinks, so that flow becomes an edge and the ranking a network property. Signed order-flow methods such as the Lee--Ready trade classification~\cite{leeready} sign trades from quotes; we treat them as future input (Section~\ref{Section:limits}).

\paragraph{Origin--destination estimation from marginals} Recovering a flow matrix from its row and column totals is the classic origin--destination problem of transport planning. The gravity model and its entropy-maximizing derivation~\cite{wilson} give the most probable matrix consistent with the marginals, $F_{ij}\propto O_i D_j$, optionally modulated by an impedance or affinity term. Section~\ref{Section:flows} is a singly constrained gravity model of this kind, with return correlation as the affinity.

\paragraph{Communities and spectral partitioning} Louvain~\cite{louvain} greedily maximizes modularity in two alternating phases (local moves, then aggregation). Recursive spectral bisection (RSB)~\cite{rsb} splits a graph by the sign or median of the Fiedler vector~\cite{fiedler}, the second eigenvector of the graph Laplacian; we use the normalized Laplacian~\cite{chung} of the community graph and the generalized Hilbert curve~\cite{hilbert,gilbert} to turn the spectral order into contiguous regions of a lattice.

\paragraph{Scattered-data interpolation and smoothing} Shepard's inverse-distance weighting (IDW)~\cite{shepard} interpolates scattered values with weights $d^{-q}$. Centroidal Voronoi tessellations and Lloyd's algorithm~\cite{cvt,lloyd} relax generators toward the density-weighted centroids of their cells; our display smoother borrows that density weighting. Gouraud shading~\cite{gouraud} interpolates vertex colours across triangles.

\paragraph{Double-link adjacency} The double link structure (DLS)~\cite{dls} keeps both directions of mesh adjacency so that upward and downward queries are equally cheap. The \MR\ solver keeps the transition matrix and its transpose for the same reason (Section~\ref{Section:engine}).

\section{Model}
\label{Section:model}

\paragraph{Paired flow matrix} For $N$ active stocks, $T\in\mathbb{R}_{\ge 0}^{N\times N}$ holds $T_{j\to i}$, the dollars that left $j$ (net selling) and entered $i$ (net buying) over the data window, with no diagonal. Section~\ref{Section:flows} describes how $T$ is estimated.

\paragraph{Out-share functor} Row-normalizing $T$ gives the transition matrix $P_{j i}=T_{j\to i}/\sum_k T_{j\to k}$, and the functor $P_i(r_j)=r_j P_{ji}$ is the part of $j$'s rank that $j$ passes to $i$. $P$ is the transition matrix of a finite Markov chain on the stocks, and Equation~(\ref{eqn:mrc}) is the stationary condition $\pi=(1-p)\,\pi P+(p/N)\mathbf{1}$ of its damped version. The solution $r=\pi$ is the vector of stationary probabilities.

\paragraph{Damping and dangling stocks} We use $p=0.15$ ($\alpha=1-p=0.85$), solved on the active sub-index. A stock with no kept outflow is a \emph{dangling} node: its row is empty, and its mass $d$ is spread uniformly, as in PageRank, so that
\begin{equation}
\pi'=\alpha\left(\pi P+\frac{d}{N}\mathbf{1}\right)+\frac{1-\alpha}{N}\mathbf{1}.
\end{equation}
Power iteration starts from the uniform vector (warm-started from the previous bar in the pipeline) and stops when $\lVert\Delta\pi\rVert_1<10^{-10}$ or after 1000 iterations. The \emph{MarketRank score} is $\pi_i N$ (1 = average), and the \emph{heartbeat} is the pulse of the score, $\Delta\log(\pi_i N)$ between consecutive bars. Using the score rather than $\pi_i$ removes the common-mode term $-\Delta\log N$ that a change in the number of active stocks adds to every $\log\pi_i$, so a stock whose score did not move reads 0.

\begin{figure}[t]
\centering
\begin{tikzpicture}[>={Stealth[length=2.2mm]},
  st/.style={circle,draw,thick,minimum size=9mm,font=\bfseries},
  fl/.style={->,thick,bend left=14},
  lb/.style={font=\scriptsize,fill=white,inner sep=1pt}]
  \node[st] (A) at (0,0) {A};
  \node[st] (B) at (5.2,0) {B};
  \node[st] (C) at (2.6,3.6) {C};
  \draw[fl] (A) to node[lb]{500K} (B);
  \draw[fl] (B) to node[lb]{200K} (A);
  \draw[fl] (A) to node[lb]{100K} (C);
  \draw[fl] (C) to node[lb]{750K} (A);
  \draw[fl] (B) to node[lb]{1M} (C);
  \draw[fl] (C) to node[lb]{400K} (B);
\end{tikzpicture}
\caption{Worked example: three stocks and six dollar flows $T_{j\to i}$ (arrow from seller $j$ to buyer $i$).}
\label{Figure:abc}
\end{figure}
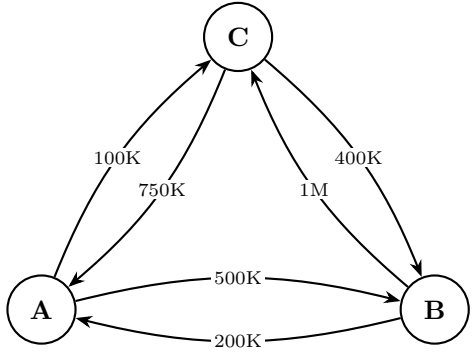

\paragraph{Worked example} Figure~\ref{Figure:abc} shows three stocks with six flows. Row-normalizing gives
\[
P=\begin{pmatrix}0 & 5/6 & 1/6\\ 1/6 & 0 & 5/6\\ 15/23 & 8/23 & 0\end{pmatrix}
\]
(rows and columns A, B, C). Table~\ref{Table:abc} lists the ranks after seven damped iterations from the uniform start and at convergence. B ranks first although C receives the most dollars (1.1M, against 0.9M for B): the chain counts shares, not dollars. B receives 5/6 of A's outflow and 8/23 of C's, and A, which gives most of its outflow to B, is itself fed by C's largest edge (15/23 of C's outflow). The unit test \texttt{test\_market\_rank.cpp} reproduces both rows through the solver (within $5\cdot10^{-5}$ after seven iterations and $10^{-5}$ at convergence) and through the pipeline's own accumulator and transition builder under the \MR\ preset.

\begin{table}[t]
\centering
\caption{Ranks of the worked example (Figure~\ref{Figure:abc}), $p=0.15$, uniform start.}
\label{Table:abc}
\begin{tabular}{lccc}
\toprule
 & B & C & A\\
\midrule
After 7 iterations & 0.35980 & 0.34770 & 0.29248\\
Converged & 0.36016 & 0.34665 & 0.29319\\
\bottomrule
\end{tabular}
\end{table}

\section{Estimating flows from public market bars}
\label{Section:flows}

Only bars are public: for each stock and bar $t$, the close $C$, the volume $V$ and the volume-weighted average price $\mathrm{VWAP}$. The estimator below is deterministic and has no fitted parameters.

\paragraph{Pressure} With return $r_i=C_i/C_i^{-}-1$, the \MR\ preset uses dollar pressure (we write $\phi$ to avoid a clash with the damping $p$)
\begin{equation}
\phi_i=r_i\,V_i\,\mathrm{VWAP}_i .
\end{equation}
A stock with $\phi_i>0$ is a net buyer, a \emph{sink}; one with $\phi_i<0$ is a net seller, a \emph{source}. Missing data gives $\phi=0$. A stock is active only while its trailing 20-bar median dollar volume is at least \$1M and its last close is at most 5 bars old. (The engine also offers $r\sqrt{V\,\mathrm{VWAP}}$ and a volume-relative pressure; they are evaluated in Section~\ref{Section:results}.)

\paragraph{Gravity pairing} Each source $i$ distributes exactly its own outflow $|\phi_i|$ over the sinks, in proportion to their pressure and tilted toward stocks it moves with:
\begin{equation}
F_{ij}=|\phi_i|\,\frac{\phi_j\,a_{ij}}{\sum_{k\in\mathrm{sinks}}\phi_k\,a_{ik}},\qquad a_{ij}=1+\lambda\rho_{ij},
\label{eqn:gravity}
\end{equation}
where $\rho_{ij}$ is the Pearson correlation of returns over the last $W=60$ bars and $\lambda=1$, so $a_{ij}\in[0,2]$. This is a singly constrained gravity model~\cite{wilson} with correlation as the affinity. Writing $\rho_{ij}=u_i\cdot u_j$ for unit-length centred return vectors $u$, the denominator is $\sum_k\phi_k+\lambda\,u_i\cdot\sum_k\phi_k u_k$, so every normalizer and the exact row and column totals cost $O(NW)$ and no $N\times N$ matrix is ever formed.

\paragraph{Sparsity} Each source stores edges only to its top 64 sinks by $\phi_j a_{ij}$, chosen from the 256 sinks with the largest $\phi_j$, with their exact shares; the row and column totals stay exact. Per bar the flux therefore has at most $64$ edges per source.

\paragraph{Accumulation and pruning} Bar flows are accumulated per row as sorted edge lists with exponential decay, $F\leftarrow 2^{-1/h}F+f$, each row capped at 256 edges by weight. Under the \MR\ preset the half-life is $h=10^9$ bars, so $T$ is effectively the \emph{cumulative} flow over the data window. Before solving, two-sided pruning keeps the union of each row's top $k_{\mathrm{out}}=20$ and each column's top $k_{\mathrm{in}}=10$ edges (ties to the lower index); a dense chain is infeasible at 10,000 nodes. The kept edges, row-normalized, are $P$.

\paragraph{What the estimate implies} Equation~(\ref{eqn:gravity}) is a modelling assumption, not an observation. It assumes that the money a stock loses in a bar goes to stocks that gained in the same bar, preferentially to correlated ones. It cannot see a seller who parks cash, a buyer funded from outside the universe, flows that pair across days, or the many trades that net out within a bar. Large, volatile, liquid names produce the largest $|\phi|$ and therefore dominate both sides of $T$. Any rank computed from $T$ is a rank of this estimate.

\section{Engine}
\label{Section:engine}

\paragraph{Language and determinism} The engine is a single C++20 binary using OpenMP~\cite{openmp} for its parallel loops. Its determinism rule is that every parallel loop writes only to the entries it owns by index, and every reduction that crosses entries is done serially in index order (for example the sums $\sum_i a_i$ and $\sum_i a_i u_i$ of the flux normalizer). Random numbers in tests use a portable Box--Muller transform rather than \texttt{std::normal\_distribution}. As a result the ranks, communities and rasters are bit-identical for any thread count, which the test suite checks by comparing 1-thread and 8-thread runs.

\paragraph{Forward and backward adjacency} The transition matrix is stored in compressed sparse row (CSR) form together with its transpose. The CSR answers ``where does $j$'s money go'' and the transpose ``where does $i$'s money come from''. The power iteration $y=xP$ is computed from $P^{\mathsf T}$ as a parallel \emph{pull}: each output entry $y_i$ is owned by one thread and sums over $i$'s in-edges, so it needs no atomics. Keeping both directions of adjacency is the idea of the double link structure~\cite{dls}.

\paragraph{Data lake} Bars come from the Alpaca Market Data API v2~\cite{alpaca} with the consolidated SIP feed (which on the free plan is available beyond a 15-minute delay), rate-limited below the provider's limit. They are stored in a Hive-partitioned Parquet~\cite{parquet} lake (\texttt{tf=/year=/month=}) managed by an embedded DuckDB~\cite{duckdb} catalogue: writes are batched with a sequence number and published by atomic rename, reads keep the newest version of each bar, and re-adjusted histories (splits, dividends) are detected by comparing re-fetched closes and refetched in full.

\paragraph{Universe} The snapshot universe is every tradable US-listed stock on Alpaca after dropping warrants, units and rights, ranked by median daily dollar volume over 20 days, keeping the top 10,000. ETFs stay as ordinary nodes (Section~\ref{Section:etf}). Sectors come from S\&P 500 GICS where known and SEC EDGAR~\cite{edgar} SIC codes otherwise; after that fill, 95.4\% of the 10,000 tickers are classified, and 4,380 of them are ETF/Fund. Every per-bar data structure is $O(Nk)$ or $O(NW)$; none is $O(N^2)$.

\paragraph{Runtime} On the real 10,000-ticker replay (250 daily bars, 22 threads), one core step of the \MR\ preset (flux, pruning, solve, forecast) takes 146.5\,ms, and the evaluation harness reports 76--282\,ms per step across configurations. The full 16-configuration evaluation over 60 bars runs in 5\,min\,07\,s at a peak resident set of 1.16\,GB. Server and landscape timings are in Section~\ref{Section:results}.

\section{The \FS\ landscape}
\label{Section:landscape}

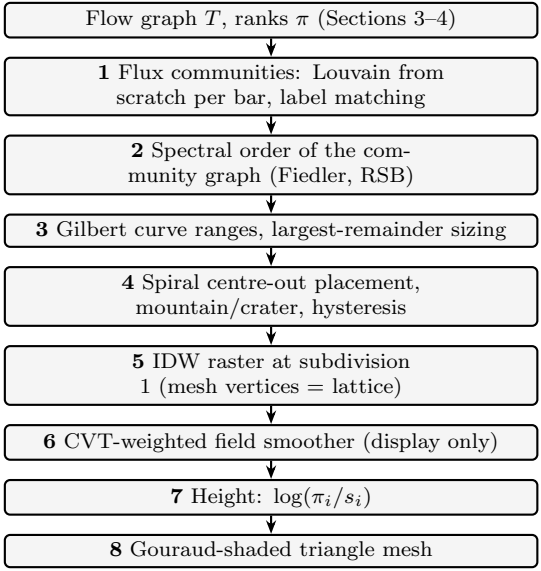
\begin{figure}[t]
\centering
\begin{tikzpicture}[node distance=2.4mm,
  box/.style={draw,rounded corners=2pt,thick,fill=gray!8,text width=0.88\columnwidth,
              align=center,font=\footnotesize,inner sep=2.5pt},
  ar/.style={-{Stealth[length=1.8mm]},thick}]
  \node[box] (f) {Flow graph $T$, ranks $\pi$ (Sections~\ref{Section:model}--\ref{Section:flows})};
  \node[box,below=of f] (s1) {\textbf{1} Flux communities: Louvain from scratch per bar, label matching};
  \node[box,below=of s1] (s2) {\textbf{2} Spectral order of the community graph (Fiedler, RSB)};
  \node[box,below=of s2] (s3) {\textbf{3} Gilbert curve ranges, largest-remainder sizing};
  \node[box,below=of s3] (s4) {\textbf{4} Spiral centre-out placement, mountain/crater, hysteresis};
  \node[box,below=of s4] (s5) {\textbf{5} IDW raster at subdivision 1 (mesh vertices = lattice)};
  \node[box,below=of s5] (s6) {\textbf{6} CVT-weighted field smoother (display only)};
  \node[box,below=of s6] (s7) {\textbf{7} Height: $\log(\pi_i/s_i)$};
  \node[box,below=of s7] (s8) {\textbf{8} Gouraud-shaded triangle mesh};
  \foreach \a/\b in {f/s1,s1/s2,s2/s3,s3/s4,s4/s5,s5/s6,s6/s7,s7/s8} \draw[ar] (\a) -- (\b);
\end{tikzpicture}
\caption{The \FS\ pipeline, in the order it runs per frame. Steps 1--4 are placement, steps 5--8 display; display-only changes redraw cached frames without re-running the model. Step 7 sets the vertex values that steps 5--6 interpolate; it is listed where it is defined in the text.}
\label{Figure:pipeline}
\end{figure}

\FS\ maps each frame to a terrain on an $n\times m$ lattice with $n=\lceil\sqrt{N}\rceil$ columns and $\lceil N/n\rceil$ rows (79$\times$79 for 6,188 active stocks). Neighbouring cells should hold stocks that trade money with each other, height should show where money settles, and a stock should not jump around between bars. Figure~\ref{Figure:pipeline} shows the pipeline; every step is deterministic, and nothing is $O(N^2)$.

\subsection{Flux-community territories}
\label{Section:louvain}
The community graph is built from a separate, short-memory accumulation of the same bar flows, with half-life $h_c=20$ bars (about a month), not from the cumulative $T$ that defines $\pi$: the communities should describe where money is moving now. Its off-diagonal kept edges among active stocks, weighted by raw flow $R_c$ and symmetrized, give $W=R_c+R_c^{\mathsf T}$. A two-phase Louvain~\cite{louvain} at resolution 1.0 runs deterministically: nodes are visited in ascending index, ties go to the lowest community id, at most 50 passes per level, and levels aggregate until nothing changes. Communities smaller than 8 stocks merge into their most strongly connected neighbour; those with no neighbour form one ``loose'' pool laid out last; at most 256 communities are kept.

The clustering is recomputed \emph{from scratch on every bar}: no warm start from the previous partition and no refresh cadence, so any stability in the map comes from the data and its 20-bar memory, not from the algorithm. Only the identity of communities is carried over, for colours and layout: new communities are matched greedily by overlap to old labels, accepting a pair when the Jaccard overlap is at least 0.3 or at least 60\% of the new community lies inside the old one (containment); unmatched communities get fresh labels, and matched labels keep their previous layout order. The first 5 bars only warm up the model.

\paragraph{How long do flux communities persist?} We measured it rather than assumed it. On the ten-year panel, a from-scratch Louvain partition of the stocks (ETFs and funds excluded, about 2,800 active stocks) was computed at 505 anchor bars and again $k$ bars later, and the two were compared by normalized mutual information (NMI) on the stocks active at both. Two references bound the result: the \emph{ceiling}, the agreement of two Louvain runs with different node orders on the same bar, and the \emph{chance} level, the same partition with labels shuffled at fixed community sizes. Table~\ref{Table:persist} gives the agreement across seeds as a share of the ceiling. The structure decays at about the rate its window forgets: with a 3-bar memory it is half gone within about 3 bars, with 20 bars within about 10, with 60 bars within about 20, and at every window it is near chance well before it has outlived its data. On the single-bar flows consecutive days agree at NMI 0.04, although each day's partition is well defined (ceiling 0.79). The cumulative graph that defines $\pi$ looks far more stable (NMI 0.86 one bar apart), but that is because about 95\% of its weight 60 bars ahead is already present today; its same-bar ceiling is only 0.54, and it has about 7 weak communities (modularity 0.12). In every case the communities are not sectors (NMI with the SEC sectors 0.03--0.04). Louvain takes 11--30\,ms per bar at this size, so per-bar re-clustering is cheap.

\begin{table}[!htbp]
\centering\footnotesize
\begin{tabular}{lccccc}
\hline
half-life & $k=1$ & 5 & 10 & 20 & 60\\
\hline
3 & 63\% & 13\% & 3\% & 1\% & 1\%\\
10 & 83\% & 47\% & 25\% & 8\% & 2\%\\
20 (used) & 89\% & 63\% & 44\% & 23\% & 3\%\\
60 & 95\% & 80\% & 67\% & 50\% & 18\%\\
\hline
\end{tabular}
\caption{Persistence of from-scratch flux communities: NMI between the partitions of bar $t$ and bar $t+k$ (flow half-life in bars; different Louvain node orders), as a share of the same-bar ceiling; mean over 505 anchors, 2016--2026. Chance is below 1\% of the ceiling at every window.}
\label{Table:persist}
\end{table}

\subsection{Spectral order of the community graph}
Communities that trade with each other should be adjacent on the lattice. The community graph (summed weights) is ordered by recursive balanced bisection: compute the Fiedler vector~\cite{fiedler} of the normalized Laplacian $L_{\mathrm{sym}}=I-D^{-1/2}WD^{-1/2}$~\cite{chung} by 500 power iterations on $2I-L_{\mathrm{sym}}$ with the trivial vector deflated, sort the communities by it, and split where the cumulative \emph{stock} count is nearest half, as in RSB~\cite{rsb}; disconnected sets fall back to their components. Sign alone orients each recursive call arbitrarily (a chain A--B--C--D can come out BADC), so after recursing each half is reversed if needed so that its end nearer the other half is the one more strongly connected to it (\emph{half-orientation}).

\subsection{Generalized Hilbert curve and territory sizing}
The lattice cells are ordered by the generalized Hilbert (``gilbert2d'') curve~\cite{hilbert,gilbert}, which visits every cell of an arbitrary $n\times m$ rectangle once, with consecutive cells at Chebyshev distance 1. Each community takes a consecutive range of the curve in spectral order, so it owns one contiguous region. With $A$ active stocks and $C\ge A$ cells, community $g$ with $a_g$ stocks gets $c_g=a_g$ cells plus its largest-remainder share of the $C-A$ spare cells (ties to the lower id), so the areas are proportional to the stock counts and sum exactly to $C$.

\subsection{Spiral placement, mountains and craters, hysteresis}
Inside a territory the cells form a spiral: they are sorted by ring $\lfloor\sqrt{d^2}\rfloor$ around the territory centre, then by angle $\mathrm{atan2}(dy,dx)$ from $-\pi$, then by index, so consecutive slots are neighbours. Stocks are ranked by a smoothed value $s=(1-\beta)v+\beta s_{\mathrm{prev}}$ ($\beta=0.5$) of the displayed quantity $v$ and take the slots in rank order. A territory is a \emph{mountain} if its median $s$ is at or above the median over all active stocks, and a \emph{crater} otherwise; a mountain puts its highest values nearest the centre, a crater its lowest. The outer $c_g-a_g$ slots stay empty and form seams between territories.

\emph{Hysteresis}: a stock keeps its previous cell if it was active last frame, is in the same community, the lattice size is unchanged, that cell still lies inside its territory, and the cell's slot in the current spiral is within $\max(2,\,0.15\,c_g)$ slots of its new slot. Conflicts resolve in rank order: kept cells first, then the ideal slot if free, then the nearest free slot of the territory.

\subsection{IDW at subdivision 1}
The raster has $(n\,s)\times(m\,s)$ pixels with subdivision $s=1$ by default, so the raster vertices \emph{are} the lattice points and become the mesh vertices. Each pixel is the Shepard~\cite{shepard} interpolant $z=\sum_i w_i h_i/\sum_i w_i$ with $w_i=d_i^{-2}$ over occupied cells within radius 3 cells; a pixel on a stock takes its value exactly, and a pixel with no stock in range takes its nearest stock, found by one multi-source breadth-first search per frame.

\subsection{CVT-weighted field smoother}
IDW leaves a ``Mexican hat'': an isolated peak is surrounded by a ring that dips toward the background. The display smoother is a Lloyd-style relaxation~\cite{lloyd,cvt} with the pixels as generators. A pixel's Voronoi cell is its square, so its neighbourhood is the $3\times3$ block with area weights 1 (centre and edge neighbours) and 0.5 (diagonals). With density $\rho=\varepsilon+|z|$, $\varepsilon=0.1\cdot\mathrm{P90}(|z|)$, each Jacobi iteration sets
\begin{equation}
z'=(1-\lambda)\,z+\lambda\,\frac{\sum a\rho z}{\sum a\rho},
\end{equation}
with 12 iterations and $\lambda=0.6$. Tall neighbours pull a pixel toward them, which fills the ring without moving any stock. Vertices of portfolio holdings are excluded from the update and keep their exact values (Section~\ref{Section:gouraud}). On a $15\times15$ test lattice with one spike of 10 on a background of 1 and its 8-neighbour ring at 0, the ring dip (far background minus the ring minimum) is 0.358 without smoothing, $-0.031$ with a Gaussian of $\sigma=1$ cell and $-0.477$ with the CVT smoother. The negative values mean the ring ends \emph{above} the background: the smoother overshoots into a halo, and it does not conserve the mean height. The smoothing is for the eye only; the ranking table, the tooltips and the API always show exact, unsmoothed values.

\subsection{Height: \texorpdfstring{$\pi$}{pi} relative to size}
Raw $\pi$ is dominated by size, since the largest names carry the largest flows. The landscape height under the \MR\ preset is therefore $\log(\pi_i/s_i)$, where $s_i$ is stock $i$'s size share (its trailing median dollar volume over the sum over active stocks). A hill attracts more money than its size predicts, a valley less, and 0 is exactly as predicted; stocks without a size get no height. The value also drives the spiral ordering and the mountain/crater rule, so it decides placement as well as relief. The \emph{ranking} is always $\pi$ itself.

\subsection{Gouraud-shaded mesh}
\label{Section:gouraud}
Each lattice quad is split into two triangles with a fixed index buffer, so per frame only the vertex heights change. Vertex colours come from a diverging map symmetric around 0 (red for inflow, blue for outflow), scaled robustly by the 90th percentile of $|z|$ and soft-clipped by $\tanh$, and each triangle is Gouraud-shaded~\cite{gouraud} with relief lighting. The user's portfolio is drawn as screen-facing dots at the holdings' own surface vertices, depth-tested with the terrain so that they move with it. Portfolio holdings are \emph{pinned}: during CVT smoothing their vertices are held at their exact values (a Dirichlet constraint), so the surface passes exactly through each holding while every other vertex is smoothed. On the real 2026-10-01 frame the measured gap between each holding's exact value and the surface is at most $4\cdot10^{-8}$, which is float rounding.

\paragraph{A fixed interface} The web interface has no model or landscape options. Its settings are fixed defaults: the \MR\ preset, height $\log(\pi/s)$, flux-community territories, the CVT smoother (12 iterations, $\lambda=0.6$) and subdivision 1. The exact sorted $\pi$ table is the truth; the surface is for the eye and the trend. Developer options remain on the command line and the HTTP API.

\subsection{The view: sections, paths and the chain around a stock}
\label{Section:view}
The interface (Figure~\ref{Figure:landscape}) is built around the chain rather than the picture. The default \emph{Sections} view shows the landscape from above with a crosshair; two profile charts cut the surface along the crosshair's row and column and plot every stock on the cut at its exact value, so a hill can be read stock by stock. Orthographic top, front and side views and a perspective 3D view are one click away, and every panel can be expanded to the full screen. The exact ranked table stays beside the landscape at all times.

Any stock can be \emph{tracked}, from the table or by typing its ticker. Its path over the whole cached history is drawn as $\log(\pi/s)$ per bar with a cursor at the displayed bar; dragging the chart scrubs the whole view through time, and several stocks share one chart. A strip under the path colours each bar by the stock's flux community, which makes the churn of Table~\ref{Table:persist} visible stock by stock.

\begin{figure}[!htbp]
\centering
\includegraphics[width=0.85\linewidth,trim=200 190 230 205,clip]{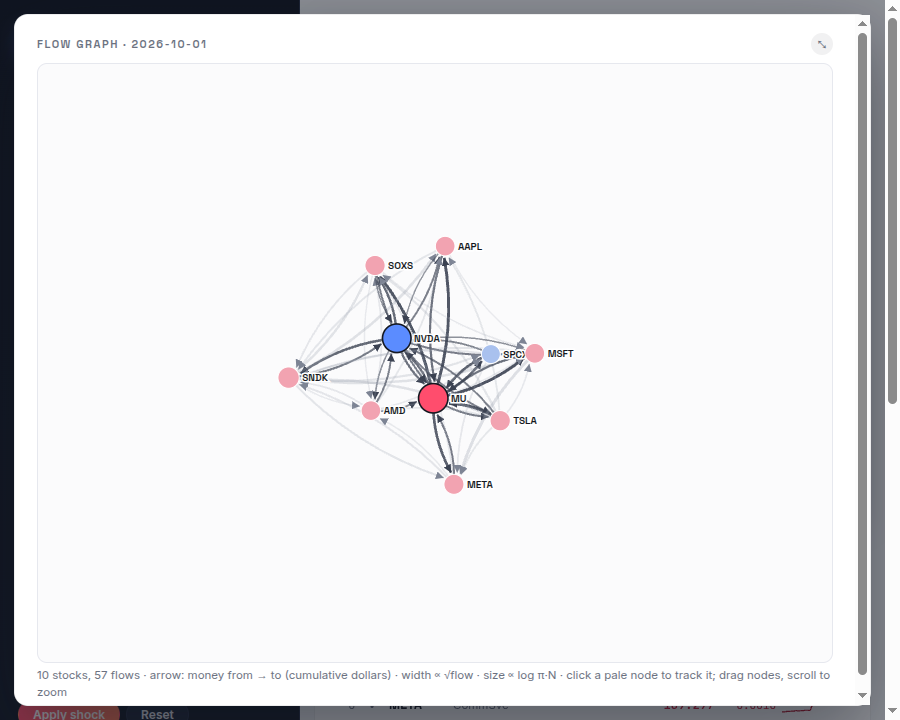}
\caption{The transition graph around MU and NVDA (large nodes) on 2026-10-01: their six strongest flow partners each and the 57 kept edges among the ten stocks. Arrows point from the stock money leaves to the stock it reaches; width grows with cumulative dollars, node size with $\log\pi N$. Dark edges touch a tracked stock; pale edges are context among the partners.}
\label{Figure:flowgraph}
\end{figure}

The \emph{transition graph} of the tracked stocks is the chain itself, seen locally (Figure~\ref{Figure:flowgraph}). For every cached bar the server keeps each active stock's six strongest kept out-edges and six strongest in-edges of $P$; for the tracked stocks it returns them, their strongest partners and every stored edge among those partners. An arrow $j\to i$ is money that left $j$ for $i$; its width grows with the cumulative dollars, and its hover gives the transition probability $P_{ji}$, the share of $j$'s kept outflow that goes to $i$, which is exactly the term of Equation~(\ref{eqn:mrc}) through which $j$ passes its rank to $i$. The layout is a force-directed simulation (d3-force~\cite{d3}) whose node positions persist from bar to bar, so the local chain visibly re-wires as the displayed bar moves.

\section{Results on real data}
\label{Section:results}

\paragraph{Data} The universe is the 10,000-ticker liquidity snapshot with 250 daily (1d) bars from the lake, replayed offline. After the SEC EDGAR fill, 95.4\% of tickers have a sector: 4,380 are ETF/Fund, 1,031 Financials, 961 Health Care, 808 Industrials and 641 Information Technology, and only 464 (4.6\%) are Unclassified. On the latest bar, 2026-10-01, 6,188 stocks are active under the \$1M liquidity floor. All measurements below use the \MR\ preset unless stated otherwise.

\paragraph{ETFs: a design position}
\label{Section:etf}
ETFs are deliberately kept in the universe and in the graph, uncapped and unfiltered, including leveraged and inverse products. In a liquidity-ranked universe they are about 44\% of the names. Where they dominate the flow, that \emph{is} the trend, and portfolios can hold ETFs as well as stocks. Leveraged and inverse products rank high because they are real dollar destinations: inflow to an inverse semiconductor ETF is a bearish signal on semiconductors.

\paragraph{Top of the ranking} Table~\ref{Table:top} lists the ten highest MarketRank scores $\pi N$ with their heartbeat $\Delta\log(\pi N)$. The top is the large-cap semiconductor, memory and technology complex, together with TSLA, META and an inverse semiconductor ETF (SOXS). Among the top 50, 26 are Information Technology and 8 are exchange-traded products: SOXS (\#5), SOXL (\#11) and SQQQ (\#15), all 3$\times$ leveraged or inverse; QQQ and SPY, with $\pi N$ of about 38--40; and the commodity trusts SLV, USO and GLD, which the SEC SIC code 6221 places in Financials rather than ETF/Fund. The leveraged semiconductor products therefore outrank the broad index funds by a wide margin. This is what the model should produce: the names that send and receive the most dollars, and receive a large share of what their large counterparties pass on, accumulate the most probability.

\begin{table}[t]
\centering
\caption{Top 10 by MarketRank score $\pi N$ (1 = average) on 2026-10-01, replay, 6,188 active stocks, SEC/GICS sectors. The heartbeat $\Delta\log(\pi N)$ is against the previous bar.}
\label{Table:top}
\small\setlength{\tabcolsep}{3.5pt}
\begin{tabular}{rllrr}
\toprule
\# & Ticker & Sector & $\pi N$ & $\Delta\log(\pi N)$\\
\midrule
1 & MU & Inf.\ Technology & 387.43 & $+0.01009$\\
2 & SNDK & Inf.\ Technology & 373.25 & $+0.00320$\\
3 & NVDA & Inf.\ Technology & 245.87 & $+0.00208$\\
4 & TSLA & Cons.\ Discretionary & 214.61 & $-0.00062$\\
5 & SOXS & ETF/Fund & 203.70 & $+0.00040$\\
6 & META & Comm.\ Services & 188.95 & $-0.0009$\\
7 & MSFT & Inf.\ Technology & 157.76 & $+0.0007$\\
8 & AAPL & Inf.\ Technology & 153.02 & $+0.0012$\\
9 & AMD & Inf.\ Technology & 151.37 & $-0.0003$\\
10 & INTC & Inf.\ Technology & 123.40 & $-0.0020$\\
\bottomrule
\end{tabular}
\end{table}

\begin{figure}[t]
\centering
\begin{tikzpicture}[>={Stealth[length=1.9mm]},
  st/.style={draw,thick,rounded corners=2pt,fill=gray!8,align=center,font=\scriptsize,inner sep=2pt},
  fl/.style={->,thick,draw=black!65,bend left=16},
  lb/.style={font=\tiny,fill=white,inner sep=0.6pt}]
  \node[st,fill=red!12] (MU) at (0,0) {\textbf{MU}\\0.333 $\cdot$ 387.4};
  \node[st] (MSFT) at (90:2.75) {\textbf{MSFT}\\0.157 $\cdot$ 157.8};
  \node[st] (AAPL) at (162:2.75) {\textbf{AAPL}\\0.130 $\cdot$ 153.0};
  \node[st] (TSLA) at (234:2.75) {\textbf{TSLA}\\0.114 $\cdot$ 214.6};
  \node[st] (SOXS) at (306:2.75) {\textbf{SOXS}\\0.152 $\cdot$ 203.7};
  \node[st] (SPCX) at (18:2.75) {\textbf{SPCX}\\0.114 $\cdot$ 105.3};
  \draw[fl] (MU) to node[lb]{2.53} (MSFT);
  \draw[fl] (MSFT) to node[lb]{1.15} (MU);
  \draw[fl] (MU) to node[lb]{2.39} (AAPL);
  \draw[fl] (AAPL) to node[lb]{1.20} (MU);
  \draw[fl] (MU) to node[lb]{0.93} (TSLA);
  \draw[fl] (TSLA) to node[lb]{2.46} (MU);
  \draw[fl] (MU) to node[lb]{2.38} (SOXS);
  \draw[fl] (SOXS) to node[lb]{0.88} (MU);
  \draw[fl] (MU) to node[lb]{1.77} (SPCX);
  \draw[fl] (SPCX) to node[lb]{1.83} (MU);
  \draw[->,thick,draw=black!65] (TSLA) to node[lb]{1.04} (SOXS);
  \draw[->,thick,draw=black!65] (TSLA) to node[lb]{0.73} (AAPL);
\end{tikzpicture}
\caption{The real graph around MU on 2026-10-01: MU and its five strongest partners, with the 12 strongest of the 30 estimated flows among them in \$B (81\% of the slice's dollars). Each box gives the slice $\pi$ (the MRC re-solved on the six stocks, $p=0.15$) and the global $\pi N$. Data: \texttt{docs/img/marketrank-slice.json}.}
\label{Figure:slice}
\end{figure}
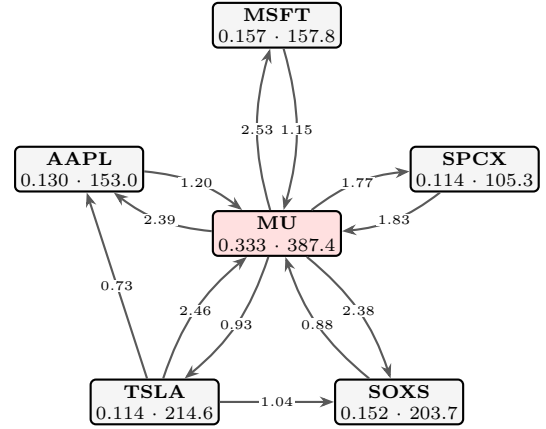

\paragraph{A slice of the real graph} Figure~\ref{Figure:slice} is the real-data counterpart of the A/B/C example (Figure~\ref{Figure:abc}). It shows the top-$\pi$ stock, MU, with its five strongest partners by raw flux in both directions, on the same bar. The arrows are estimated dollar flows accumulated over the data window, not observed trades. ``Slice $\pi$'' is the MRC re-solved on these six stocks alone with $p=0.15$. MU takes a third of the slice's probability (0.333), although within the slice it sends more (\$10.0B) than it receives (\$7.5B): the chain counts out-shares, not net dollars, and every partner sends MU the largest share of its slice outflow. TSLA is second of the six in the full graph ($\pi N$ 214.6) but last in the slice, so its global standing rests on money from stocks outside the slice.

\begin{figure*}[t]
\centering
\includegraphics[width=\textwidth,keepaspectratio]{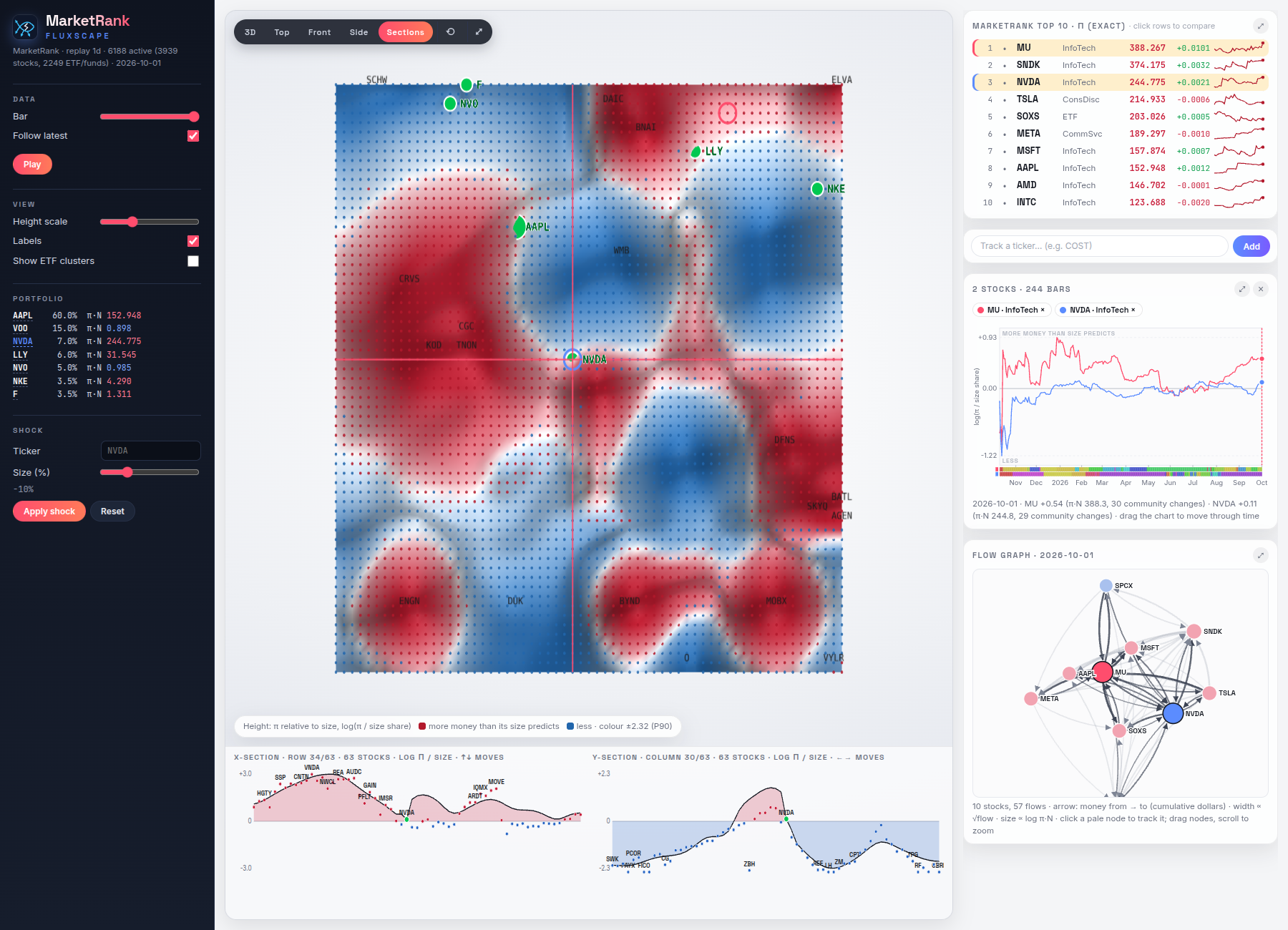}
\caption{The \MR\ \FS\ interface on the real 2026-10-01 bar (6,188 active names: 3,939 stocks and 2,249 ETF/funds, which are in the chain but hidden from the surface by default), \MR\ preset. \emph{Centre:} the default Sections view, the landscape seen from above with a crosshair; height and colour are $\pi$ relative to size, $\log(\pi/s)$, on a red/blue diverging scale at P90 $\pm2.32$, and the two profile charts below cut the surface along the crosshair's row and column with every stock on the cut. Portfolio holdings are green dots pinned to their exact values. \emph{Right, top:} the truth table, the ten highest exact $\pi N$ with the heartbeat $\Delta\log(\pi N)$. \emph{Right, middle:} the tracked stocks (MU, NVDA) over the whole cached year, as $\log(\pi/s)$ per bar with a strip of each stock's flux community and a cursor at the displayed bar. \emph{Right, bottom:} the transition graph around the tracked stocks (Figure~\ref{Figure:flowgraph}).}
\label{Figure:landscape}
\end{figure*}

\paragraph{Contrast with hotness} The engine's earlier ``money-flow'' preset reads a different quantity, hotness relative to size, $h_i=\pi_i/\pi^{\mathrm{ref}}_i-1$ with $\pi^{\mathrm{ref}}$ proportional to trailing median dollar volume, on a chain in which net buyers retain part of their inflow. Its top 10 on the same data are TNON, KOD, DFNS, NFEGP, SOAR, BNAI, WETO, LHSW, AGEN and GRML: all small names. Dividing by size rewards any stock whose flow is large relative to a small denominator, so a single volume spike in an illiquid name outranks steady flow in a mega-cap. The ratio is informative, but as a primary ranking it is dominated by noise in the smallest names, which is why \MR\ ranks by $\pi$ and uses size only for the landscape height.

\paragraph{Evaluation} The harness runs the pipeline over the last 60 of 250 daily bars (10,000 nodes) for 16 configurations. For each configuration it reports:
\begin{itemize}
\item the \emph{floor share}, the fraction of active stocks at the teleport floor;
\item the Gini coefficient of $\pi$;
\item \emph{sector coherence}, defined below;
\item \emph{structure gain}, $1-$ Spearman of $\pi$ against inflow share (near 0 means the chain has collapsed to plain inflow);
\item the information coefficient (IC), the mean daily Spearman correlation with the next bar's return, with $t=\bar{x}/(s/\sqrt{n})$, for the $+1$-bar forecast score and for hotness $h$.
\end{itemize}
Table~\ref{Table:eval} gives the result. ``Legacy'' is the first milestone (dollar pressure, no inbound pruning, no retention). +A to +E switch on, one at a time, relative pressure, excess-over-gravity lift, $k_{\mathrm{in}}=10$, size or long-run hotness references, and retention. ``Defaults'' has them all, and ``money-flow'' is dollar pressure with two-sided pruning and retention.

\begin{table*}[t]
\centering
\caption{Evaluation grid on real data: last 60 of 250 daily bars, 10,000 nodes, 22 threads, SEC sectors. IC is the mean daily Spearman correlation with the next bar's close-to-close return; $t$ its $t$-statistic. Coherence is the engine metric (ETF/Fund counted as a sector; see Table~\ref{Table:coh}). The last column is the mean core step time.}
\label{Table:eval}
\small\setlength{\tabcolsep}{4pt}
\begin{tabular}{lrrrrrrrrr}
\toprule
Configuration & Floor & Gini & Coherence & Structure & IC(score) & $t$ & IC($h$) & $t$ & ms/frame\\
\midrule
legacy             & 0.947 & 0.847 & 0.226 & 0.642 & $+0.0093$ & $+1.95$ & $-0.0240$ & $-3.95$ & 181.0\\
+A relative        & 0.949 & 0.843 & 0.156 & 0.643 & $+0.0144$ & $+5.04$ & $-0.0471$ & $-11.03$ & 187.1\\
+B excess          & 0.929 & 0.843 & 0.207 & 0.591 & $+0.0095$ & $+1.81$ & $-0.0267$ & $-4.27$ & 116.2\\
+C $k_{\mathrm{in}}=10$ & 0.850 & 0.840 & 0.226 & 0.419 & $+0.0098$ & $+1.71$ & $-0.0274$ & $-4.30$ & 244.2\\
+D size            & 0.947 & 0.847 & 0.226 & 0.642 & $+0.0093$ & $+1.95$ & $-0.0043$ & $-0.53$ & 187.0\\
+D longrun         & 0.947 & 0.847 & 0.226 & 0.642 & $+0.0093$ & $+1.95$ & $+0.0032$ & $+0.84$ & 282.1\\
+E retention       & 0.001 & 0.760 & 0.226 & 0.628 & $-0.0094$ & $-0.72$ & $-0.0243$ & $-1.18$ & 198.9\\
defaults           & 0.001 & 0.743 & 0.148 & 0.429 & $-0.0164$ & $-1.20$ & $-0.0195$ & $-1.30$ & 79.1\\
defaults relative  & 0.002 & 0.728 & 0.145 & 0.318 & $-0.0092$ & $-0.66$ & $-0.0306$ & $-3.54$ & 82.8\\
defaults+longrun   & 0.001 & 0.743 & 0.148 & 0.429 & $-0.0164$ & $-1.20$ & $-0.0041$ & $-0.28$ & 117.5\\
defaults+netflow   & 0.001 & 0.743 & 0.148 & 0.429 & $-0.0164$ & $-1.20$ & $-0.0103$ & $-0.36$ & 77.7\\
money-flow         & 0.001 & 0.732 & 0.257 & 0.508 & $-0.0155$ & $-1.10$ & $-0.0057$ & $-0.91$ & 170.6\\
money-flow+netflow & 0.001 & 0.732 & 0.257 & 0.508 & $-0.0155$ & $-1.10$ & $-0.0136$ & $-0.49$ & 172.9\\
defaults+volscale  & 0.002 & 0.742 & 0.120 & 0.369 & $-0.0203$ & $-1.61$ & $-0.0144$ & $-1.09$ & 76.3\\
money-flow+volscale& 0.002 & 0.747 & 0.180 & 0.493 & $-0.0217$ & $-1.66$ & $-0.0050$ & $-0.79$ & 164.5\\
\midrule
\textbf{marketrank} & \textbf{0.714} & \textbf{0.832} & \textbf{0.222} & \textbf{0.248} & $\mathbf{+0.0035}$ & $\mathbf{+0.54}$ & $\mathbf{-0.0151}$ & $\mathbf{-1.96}$ & \textbf{146.5}\\
\bottomrule
\end{tabular}
\end{table*}

\paragraph{Reading the evaluation} The \MR\ preset has floor share 0.714, Gini 0.832, structure gain 0.248, score IC $+0.0035$ ($t=0.54$) and hotness IC $-0.015$ ($t=-1.96$). The open-to-open variants (decide at the close, fill at the next open, exit at the following open) are similar: $+0.0057$ ($t=0.90$) and $-0.014$ ($t=-1.85$). Its score IC is approximately zero. \MR\ $\pi$ describes where money concentrates; it is \emph{not} a next-day return forecast.

The one significant pattern is a reversal in hotness under the legacy and relative-pressure configurations: names that are hot today underperform tomorrow. IC($h$) is $-0.047$ ($t=-11.0$) under +A relative, between $-0.024$ and $-0.027$ ($t\approx-4$) under legacy, +B and +C, and $-0.031$ ($t=-3.54$) under defaults with relative pressure. It survives the tradeable open-to-open timing ($t=-9.0$ for +A relative). The only significant positive score IC is +A relative ($+0.014$, $t=5.0$; open-to-open $t=4.6$), a configuration in which 95\% of stocks sit at the teleport floor. Sixty daily cross-sections are a short sample, and these $t$-statistics ignore cross-sectional dependence beyond the daily IC, so they are probably optimistic.

\paragraph{Sector coherence} The engine's coherence is the share of slow-flux edge weight, over edges between two active stocks with a known sector, that stays within a sector:
\[
\mathrm{coh}=\frac{\sum_{i\to j}\mathrm{raw}_{ij}\,[\mathrm{sec}_i=\mathrm{sec}_j]}{\sum_{i\to j}\mathrm{raw}_{ij}},
\]
averaged over the evaluation frames. Unclassified names are excluded, but ETF/Fund counts as a sector. Table~\ref{Table:coh} compares it with a null in which sector labels are permuted among the same nodes (5 seeds). Under the engine definition every configuration scores \emph{below} its shuffled baseline (\MR\ 0.222 against 0.263), except money-flow, which is level with it (0.257 against 0.252). ETF/Fund is 44\% of the nodes, so it dominates the random-label baseline, while the flux runs between ETFs and single stocks more often than random labels predict, which is plausible because a sector or index ETF trades with its constituents. With ETF/Fund and Unclassified both removed, that is, on stock-to-stock flow only, coherence is well above the null: 0.288 against 0.123 (2.35$\times$) under \MR, and 0.333 against 0.116 (2.9$\times$) under money-flow. This is the positive structural result: stock-to-stock edges are sector-enriched. An earlier figure of about 0.36 was computed when only some 5--9\% of names were classified, mostly the liquid S\&P 500, and is not comparable.

\begin{table}[t]
\centering
\caption{Sector coherence against a shuffled-label null (mean of 5 permutations), same frames as Table~\ref{Table:eval}. ``Engine'' counts ETF/Fund as a sector; ``stocks only'' drops ETF/Fund and Unclassified. Rows with the same transition matrix are merged.}
\label{Table:coh}
\footnotesize\setlength{\tabcolsep}{3pt}
\begin{tabular}{lrrrrr}
\toprule
 & \multicolumn{2}{c}{Engine} & \multicolumn{3}{c}{Stocks only}\\
\cmidrule(lr){2-3}\cmidrule(lr){4-6}
Configuration & coh & null & coh & null & ratio\\
\midrule
legacy, +D, +E      & 0.226 & 0.254 & 0.289 & 0.115 & 2.53\\
+A relative         & 0.156 & 0.238 & 0.168 & 0.143 & 1.17\\
+B excess           & 0.208 & 0.254 & 0.264 & 0.115 & 2.29\\
+C $k_{\mathrm{in}}=10$ & 0.226 & 0.254 & 0.289 & 0.115 & 2.51\\
defaults            & 0.148 & 0.236 & 0.183 & 0.130 & 1.41\\
defaults relative   & 0.145 & 0.233 & 0.176 & 0.134 & 1.31\\
money-flow          & 0.257 & 0.252 & 0.333 & 0.116 & 2.87\\
defaults+volscale   & 0.120 & 0.261 & 0.147 & 0.130 & 1.13\\
money-flow+volscale & 0.180 & 0.281 & 0.234 & 0.118 & 1.99\\
\textbf{marketrank} & \textbf{0.222} & \textbf{0.263} & \textbf{0.288} & \textbf{0.123} & \textbf{2.35}\\
\bottomrule
\end{tabular}
\end{table}

\paragraph{Flux communities are not sectors} On the last bar the 20-bar flow graph of the 3,939 placed stocks has 7 communities plus a loose pool of 4 names, with modularity 0.108. Against SEC sectors, over the 3,744 classified stocks outside the loose pool, the normalized mutual information is 0.034, against 0.004 for shuffled labels (20 seeds); the persistence study finds 0.03--0.04 on every flow window (Section~\ref{Section:louvain}). The agreement is above chance but negligible in absolute terms. Table~\ref{Table:comm} shows that each community's leading sectors stay within about ten points of the universe base rates (Financials 19.0\%, Health Care 16.7\%, Industrials 16.3\%, IT 13.0\%); the strongest tilts are Health Care at 32\% and Financials at 32\% in two of the smaller communities. Money rotates across sector lines, and the partition reflects recent flow regimes rather than industries. The territories therefore show structure that a sector map does not; we do not claim that they recover sectors.

\begin{table}[t]
\centering
\caption{Flux communities of the 20-bar flows on 2026-10-01 (ETF/funds hidden, the default). Size is the number of placed stocks; the top-2 sector shares are among each community's classified stocks (Fin.\ = Financials, Indust.\ = Industrials, IT = Information Technology).}
\label{Table:comm}
\footnotesize\setlength{\tabcolsep}{3pt}
\begin{tabular}{rrll}
\toprule
Comm. & Stocks & Top-1 & Top-2\\
\midrule
157 & 1391 & Fin.\ 27\% & Indust.\ 17\%\\
155 & 742 & Indust.\ 20\% & Health 18\%\\
146 & 610 & Health 22\% & Indust.\ 16\%\\
150 & 438 & Fin.\ 32\% & IT 20\%\\
159 & 297 & Health 21\% & IT 20\%\\
160 & 272 & Health 32\% & Fin.\ 14\%\\
154 & 185 & Health 23\% & IT 14\%\\
loose & 4 & -- & --\\
\bottomrule
\end{tabular}
\end{table}

\paragraph{Landscape} Figure~\ref{Figure:landscape} shows the interface on the real 2026-10-01 bar: the landscape in the default Sections view, the exact table, the paths of two tracked stocks and their transition graph.
\begin{figure*}[t]
\centering
\begin{subfigure}[t]{0.49\textwidth}\centering
\includegraphics[width=\linewidth,trim=15 15 15 15,clip]{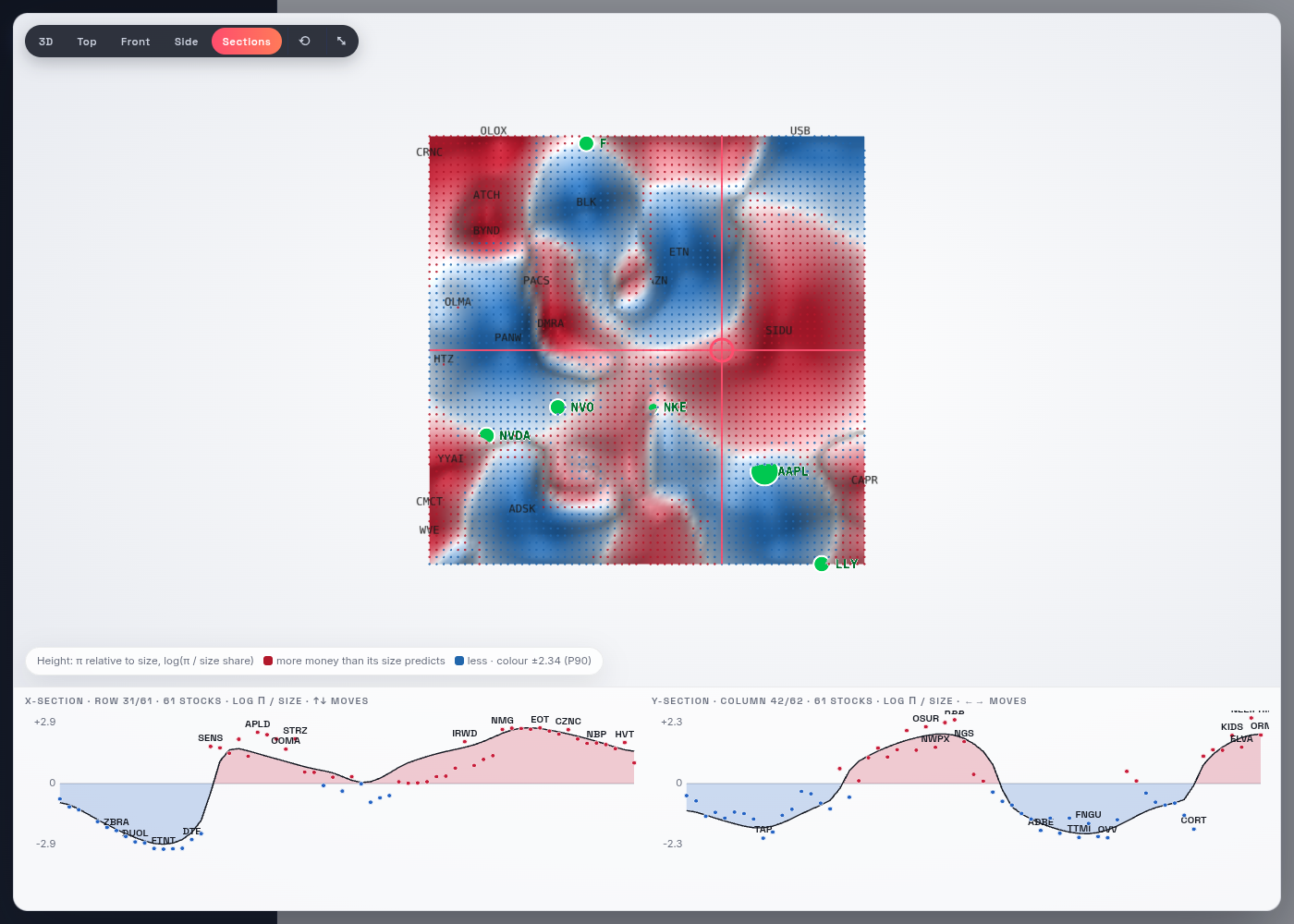}
\caption{2026-01-02}\label{Figure:tl1}
\end{subfigure}\hfill
\begin{subfigure}[t]{0.49\textwidth}\centering
\includegraphics[width=\linewidth,trim=15 15 15 15,clip]{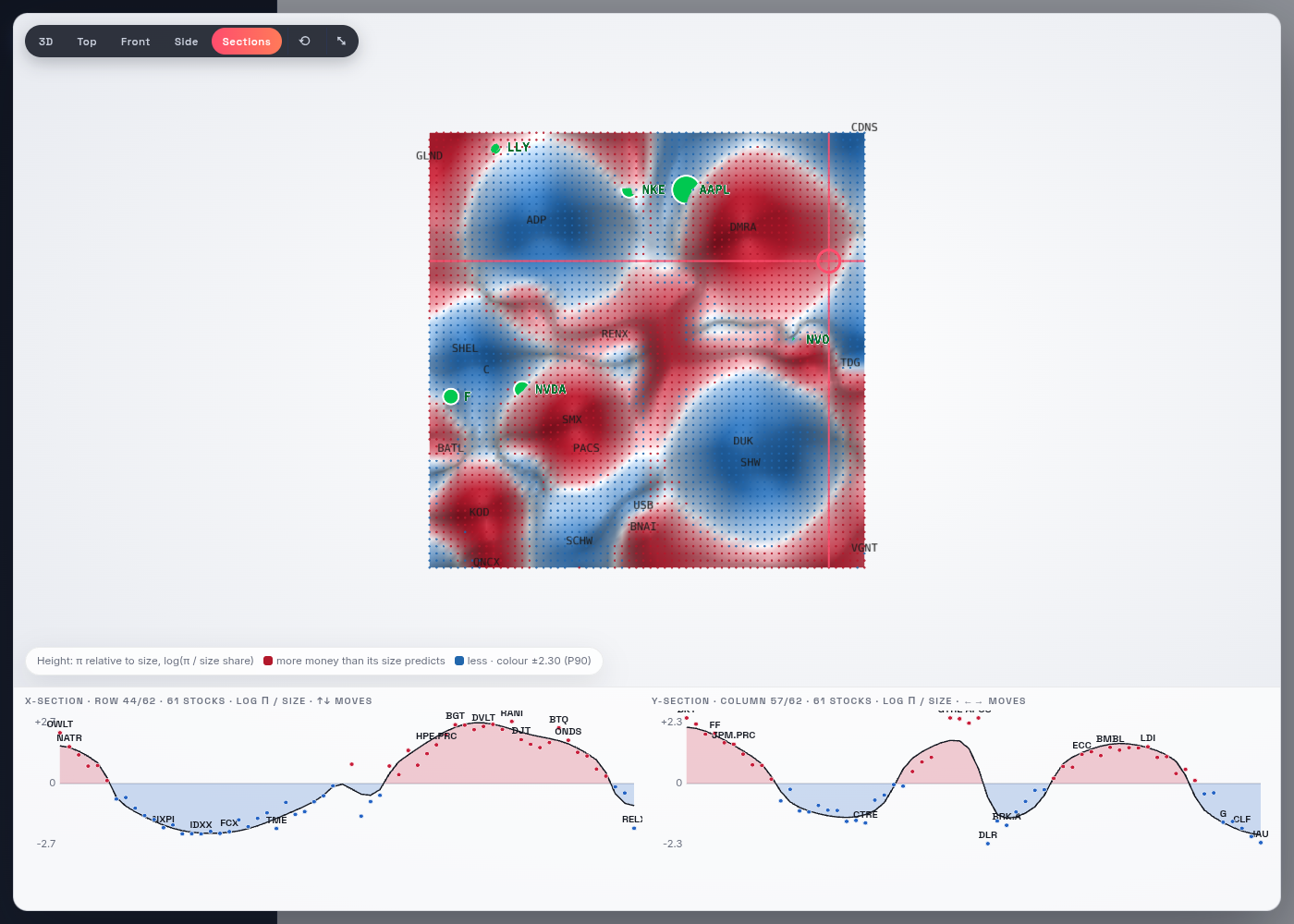}
\caption{2026-04-01}\label{Figure:tl2}
\end{subfigure}

\medskip
\begin{subfigure}[t]{0.49\textwidth}\centering
\includegraphics[width=\linewidth,trim=15 15 15 15,clip]{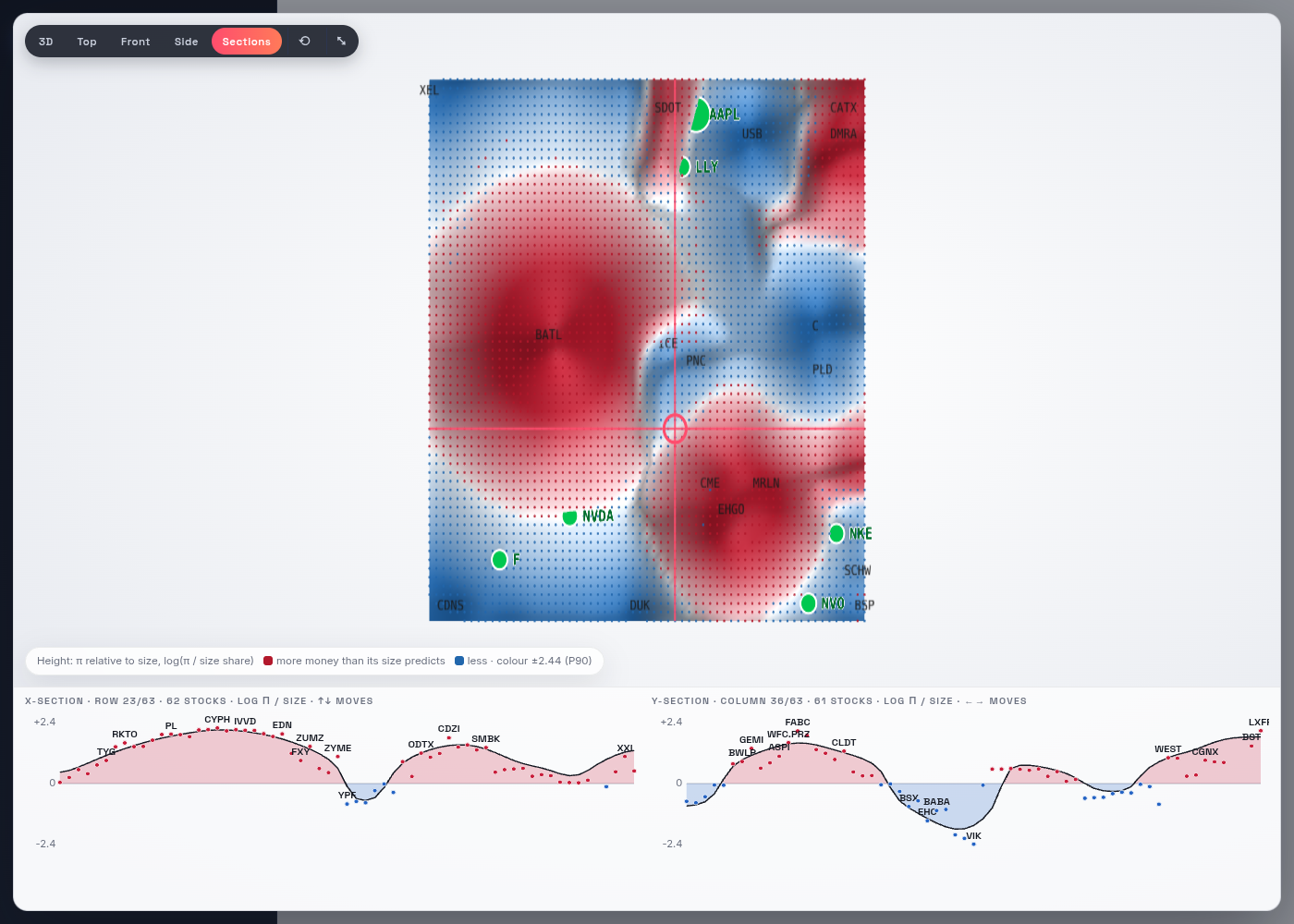}
\caption{2026-07-01}\label{Figure:tl3}
\end{subfigure}\hfill
\begin{subfigure}[t]{0.49\textwidth}\centering
\includegraphics[width=\linewidth,trim=15 15 15 15,clip]{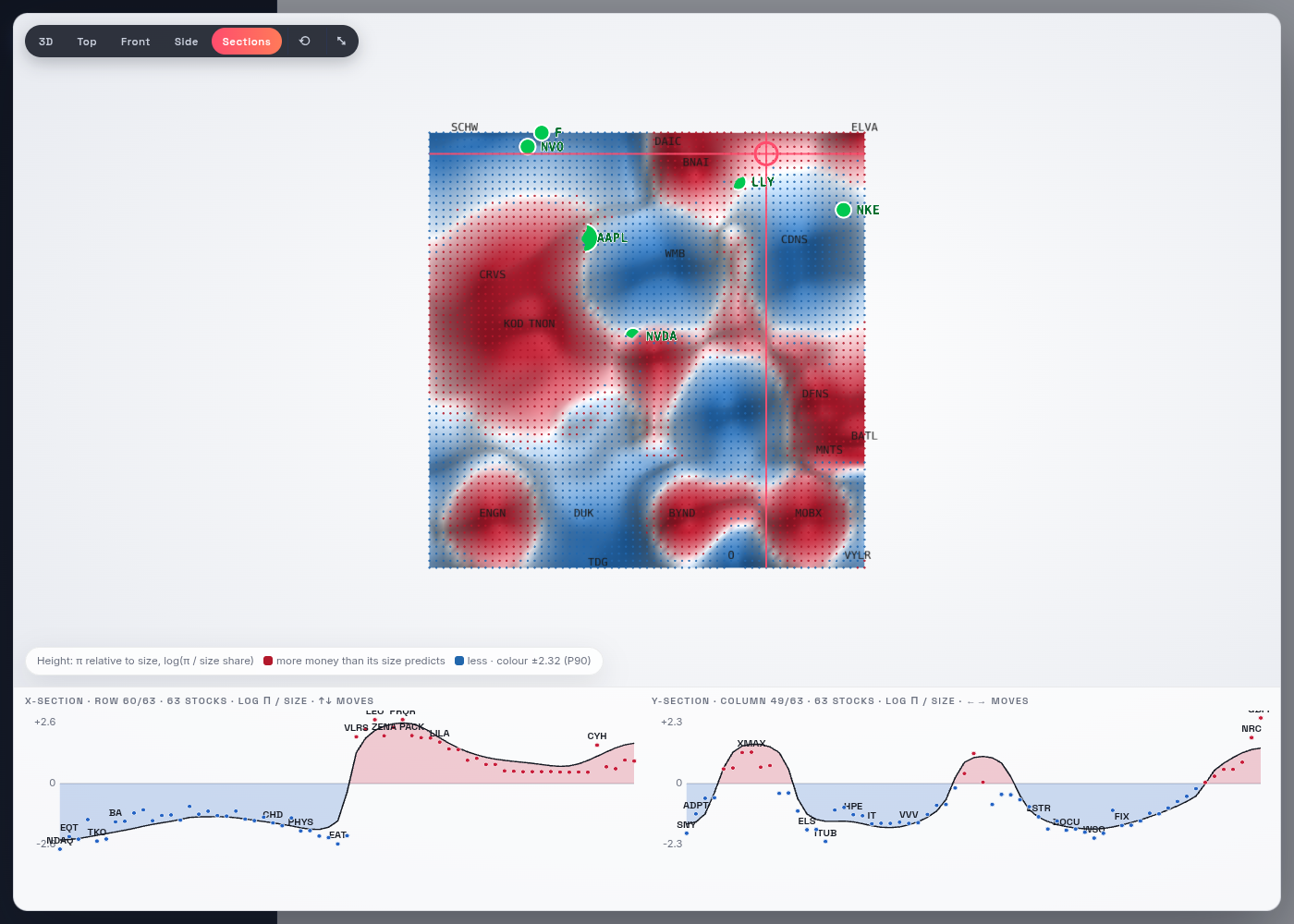}
\caption{2026-10-01}\label{Figure:tl4}
\end{subfigure}
\caption{\FS\ time-lapse on four real bars about a quarter apart, in the default Sections view expanded to the full screen: \MR\ preset, height $\log(\pi/s)$, flux communities re-clustered from scratch on each bar from the 20-bar flows, holdings pinned (green dots), with the row and column profiles under each map. Active names: 5,645, 5,879, 6,142 and 6,188.}
\label{Figure:timelapse}
\end{figure*}

\paragraph{Time-lapse} Figure~\ref{Figure:timelapse} shows the landscape on four dates about a quarter apart. Because the communities are now re-clustered from scratch on each bar from the recent flows, the territories are re-drawn from panel to panel rather than carried over: there are 7 communities in January (modularity 0.176), 7 in April (0.145), 4 in July (0.108) and 7 in October (0.108). Weakening modularity means the recent flows split less cleanly into groups as the year goes on. The ranking of the chain is far steadier than the map. The semiconductor and memory complex leads throughout while its order rotates: TSLA ($\pi N$ 455.8), NVDA, AMD, BYND and MU in January; SNDK (361.9), TSLA, NVDA and MU in April, with SLV and SQQQ close behind; MU (387.4), SNDK, TSLA, NVDA and SOXS in July; MU (388.3), SNDK, NVDA, TSLA, SOXS and META in October. The holdings stay pinned to their exact values in every panel.

\paragraph{Stability} An earlier version of the view warm-started Louvain from the previous partition and re-clustered only every 5 bars on the cumulative graph. Over 245 frames it kept 85\% of stocks in their cell between re-clusters and 74\% of community labels per re-cluster. That stability was largely produced by the algorithm and the cumulative graph, not by the market (Table~\ref{Table:persist}), and it made the map look more persistent than the flows are. The current view re-clusters from scratch on every bar on the 20-bar flows: territories now hold their shape for about a week and reorganize over a month, as the measurement predicts. The spiral ordering still keeps mountains and craters coherent within a territory (median Spearman correlation of value against distance from the centre 0.98 in the earlier measurement).

\paragraph{Performance} Measured on the server with the \MR\ preset, 10,000 nodes, 250 daily bars and 22 threads:
\begin{itemize}
\item time to ready is 42.3\,s, from process start to the status endpoint reporting ready. That includes the DuckDB load, the panel build, 249 core steps and 245 landscapes.
\item a core model step takes about 146\,ms, so the 249 sequential steps (about 36\,s) dominate start-up.
\item peak memory is 1.34\,GB: the panel plus up to 300 cached landscape frames.
\item a landscape frame takes a median of 23.1\,ms (mean 31.2\,ms).
\item a Louvain re-cluster takes a median of 54.8\,ms on the 49 re-clustering frames; only those frames pay it.
\end{itemize}
The 60-bar, 16-configuration evaluation runs in 5\,min\,07\,s.

\section{Walk-forward evaluation protocol}
\label{Section:wf}

Section~\ref{Section:results} shows that \MR\ measures structure but that its level does not forecast next-day returns. Whether its signals can nonetheless inform portfolio moves is a different question: \emph{had we followed them week by week, would the portfolio have beaten simply holding the current mix, after trading costs?} We answer it with a walk-forward protocol that was fixed, defaults and pass/fail thresholds included, before any of its results were seen. The results, reported in Section~\ref{Section:wfresults} whichever way they came out, are negative.

\paragraph{Backfill and universe} Daily SIP bars are backfilled about ten years (the backfill as run reaches 2016-06-27, 2,582 bars per ticker). Each rebalance uses a \emph{point-in-time} liquidity ranking over today's candidate list: names are ranked by trailing median dollar volume computed only from bars available up to that rebalance. Survivorship bias remains, because the data provider lists only today's tickers, so stocks delisted earlier are missing, and that favours strategies that held survivors. We report it as a limitation and re-run the whole test on a large-cap sub-universe (the point-in-time top 500), where delistings are rare. An edge that exists only in small names is treated as suspect.

\paragraph{One causal pass} The pipeline steps once through the history, and every quantity stored at a rebalance date depends only on bars up to that date. A test asserts that changing any later bar leaves every earlier decision bit-identical. The first 12 months are warm-up. At each bar the harness records eight candidate signals, each as a cross-sectional $z$-score over eligible stocks winsorized at $\pm3$:
\begin{itemize}
\item the score $\log(\pi N)$;
\item the heartbeat over 1, 5 and 20 bars, $\log(\pi N)_t-\log(\pi N)_{t-K}$;
\item $\pi$ relative to size, $\log(\pi/s)$;
\item negated hotness $-h$, motivated by the reversal of Section~\ref{Section:results};
\item inflow momentum, $\log(\mathrm{in}_t/\mathrm{in}_{t-5})$;
\item the existing forecast score.
\end{itemize}

\paragraph{Weekly as the primary test} The primary test rebalances weekly, on the last trading day of each ISO week, and executes at the next open; a monthly run (the last trading day of each month) is kept as a secondary comparison. Flow-induced pressure, if it carries information, should show within about a week; beyond that horizon, flows mix with unrelated news. Weekly trading raises turnover, which makes the cost-sensitivity runs below more important.

\paragraph{IC research table} For every signal and horizon $h\in\{1,2,5\}$ trading days, with $h=20$ kept only as a decay diagnostic, the harness reports the mean information coefficient against the open-to-open forward return (signal known at the close of $t$, entry at the open of $t+1$, exit at the open of $t+1+h$), its $t$-statistic, and its stability by calendar year. Samples are taken every $h$ bars, so they do not overlap.

\paragraph{Strategy} The base is the current mix (AAPL 60\%, VOO 15\%, NVDA 7\%, LLY 6\%, NVO 5\%, NKE 3.5\%, F 3.5\%), held fixed over history. At each weekly rebalance, a tilt of 20\% of portfolio value goes in equal weight to the top 10 eligible names by the signal, funded pro rata from the base. The rules are:
\begin{itemize}
\item long-only, with at most 10\% of portfolio value in any single tilt name;
\item a name is eligible only with a trailing median dollar volume of at least \$50M;
\item turnover is capped at 0.5 per rebalance;
\item trades execute at the next day's open, and holdings are marked at each close;
\item costs are 10\,bps per side, with sensitivity runs at 0 and 25\,bps.
\end{itemize}
The simulator also checks its own bookkeeping: after every trade, the value of what the portfolio owns plus its cash must equal what it had just before the trade, minus the trading fee. So it can neither create nor lose money through an accounting slip; any gain or loss comes only from prices and costs. The benchmarks are buy-and-hold of the base mix, the base mix rebalanced on the same weekly calendar (with costs), and VOO.

\paragraph{Gated blend} A simplex-weighted blend of the signals is learned on a trailing 156-week window with a 1-week embargo, so that no training label overlaps the week being decided. Each weight is the positive part of that signal's mean training IC, normalized to sum to 1. The blend is used only while its own out-of-sample IC over the previous 104 weeks has $t>2$, with at least 52 weeks required; otherwise the strategy holds the base with no tilt. A dead signal therefore costs nothing beyond the trades the gate avoids.

\paragraph{Metrics and multiple testing} The harness reports the annualized return, volatility and Sharpe ratio, the excess return over each benchmark, the information ratio, the maximum drawdown, turnover, cost drag and the hit rate by calendar year. The 95\% interval on annualized excess return comes from a block bootstrap with 21-day blocks. Every configuration ever run is appended to a registry with its daily Sharpe and daily information ratio, and the deflated probabilities use the registry's trial count and variance, so trying many variants is paid for rather than hidden.

\paragraph{Decision gate} The gate judges the weekly strategy. The answer is ``yes, consistently'' only if all of the following hold (spec \S3):
\begin{enumerate}
\item annualized excess return over buy-and-hold of the base mix $>0$ at 10\,bps, with a bootstrap 95\% interval above 0;
\item excess return positive in at least 60\% of calendar years;
\item a deflated probability above 0.95 for the information ratio of the daily excess over buy-and-hold (amended on 2026-10-02 after the final review; the original wording deflated the total Sharpe ratio, see Section~\ref{Section:wfresults});
\item a maximum drawdown no worse than the base mix's plus 5 percentage points;
\item the result holds on the large-cap sub-universe (survivorship check).
\end{enumerate}
Anything less is ``no''.

\subsection{Results}
\label{Section:wfresults}

The backfill covers 10,000 tickers from 2016-06-27 to 2026-10-02 (2,582 daily bars). After the 252-bar warm-up the weekly pass makes 483 rebalances, from 2017-06-30 to 2026-09-25 (2,326 trading days). Six configurations were run, all registered as trials of one registry (54 strategy rows): the main weekly run at 10\,bps, the same at 0 and 25\,bps, weekly with relative pressure (the \MR\ preset with relative pressure; not the legacy +A relative configuration of Section~\ref{Section:results}), monthly at 10\,bps, and weekly on the point-in-time top 500.

\paragraph{Information coefficients} Table~\ref{Table:wfic} gives the significant entries of the research table. One effect is persistent: the hotness reversal. Names that are cold today do slightly better over the next 1 to 20 days. The IC is $+0.006$ at one day ($t=3.2$), positive in 8 of 10 years, and it holds on the large caps ($t=3.4$). Under the \MR\ preset the score is its exact mirror. $\pi$ relative to size ranks the wrong way round, with IC $-0.009$ ($t=-4.6$) and 8 of 10 years negative. Inflow momentum is weakly negative. The heartbeats and the forecast are indistinguishable from zero. With relative pressure the reversal has $t=1.4$ at one day. This run is the \MR\ preset with relative pressure, not the legacy +A relative configuration behind the $t=-11$ of Section~\ref{Section:results}, which had $k_{\mathrm{in}}=0$, no liquidity floor, and a different universe and eligibility. Both the configuration and the sample changed, so the run neither replicates nor refutes that one-year result.

\begin{table}[t]
\centering
\caption{Walk-forward IC research table, entries with $|t|\ge2$ (weekly run, all eligible names; same panel for the large-cap run). Mean Spearman IC against the open-to-open return over $h$ bars, on non-overlapping samples; ``years'' counts calendar years with a positive mean IC; $-h$ is negated hotness.}
\label{Table:wfic}
\small\setlength{\tabcolsep}{3pt}
\begin{tabular}{llrrrr}
\toprule
Univ. & Signal & $h$ & IC & $t$ & years $+$\\
\midrule
all & $-h$ & 1 & $+0.0059$ & $+3.19$ & 8/10\\
all & $-h$ & 5 & $+0.0085$ & $+2.16$ & 7/10\\
all & $-h$ & 20 & $+0.0168$ & $+2.32$ & 7/10\\
all & $\log(\pi N)$ & 1 & $-0.0059$ & $-3.19$ & 2/10\\
all & $\log(\pi/s)$ & 1 & $-0.0094$ & $-4.55$ & 2/10\\
all & $\log(\pi/s)$ & 5 & $-0.0146$ & $-3.26$ & 2/10\\
all & inflow mom. & 5 & $-0.0140$ & $-2.38$ & 1/10\\
top 500 & $-h$ & 1 & $+0.0071$ & $+3.42$ & 8/10\\
top 500 & $\log(\pi/s)$ & 1 & $-0.0086$ & $-3.86$ & 1/10\\
relative & $-h$ & 1 & $+0.0041$ & $+1.39$ & 6/10\\
relative & inflow mom. & 1 & $-0.0078$ & $-2.87$ & 2/10\\
\bottomrule
\end{tabular}
\end{table}

\paragraph{Portfolio results} Table~\ref{Table:wfperf} compares the gated blend of every run with buy-and-hold of the base, which returned 31.2\% a year with a 32.2\% maximum drawdown. Every run is negative, and in every run the 95\% interval lies below zero. The gate of the blend opened in only 16 of 483 weeks (3.3\%): its out-of-sample IC rarely reached $t>2$. It opened in 2 weeks (0.4\%) under relative pressure, 48 (9.9\%) on the large caps and never in the monthly run. Most of the time the blend therefore held the base rebalanced weekly. That rebalancing trims AAPL and NVDA during their run, and it alone costs 3.5\% a year against buy-and-hold. Against the rebalanced base, the blend loses 0.6\% a year. The best single signal, the score (the largest flow sinks), gains 1.5\% a year ($t\approx1.3$, not significant), and the same signal loses 3.5\% a year on the large caps. Every other signal loses 3--10\% a year, and the fast heartbeats turn over about 0.2 per weekly rebalance: the whole 20\% tilt is replaced about once a week. The reversal's IC is real but small: the top 10 names by $-h$ lose 2.9\% a year against the rebalanced base. Costs do not change the conclusion; the 0 and 25\,bps runs bracket the main one within half a point.

\begin{table*}[t]
\centering
\caption{Gated blend against buy-and-hold of the base mix, 2017-07 to 2026-10. Ann.\ excess is the mean daily excess return times 252; it differs from the gap between the annualized returns. The 95\% interval is a 21-day block bootstrap; ``years $+$'' counts calendar years with positive excess. DSR$_e$ is the deflated probability of the daily information ratio of the excess over the 54 registered trials, the quantity gate c3 tests; every entry fails c3.\textsuperscript{a}\\
\textsuperscript{a}\,The deflated probability of the \emph{total} Sharpe ratio, which the gate used before the 2026-10-02 amendment, is 0.987--0.992 for all six runs. It is high because the base is, not because of the tilt, and it is kept in the reports as information only.}
\label{Table:wfperf}
\small\setlength{\tabcolsep}{4pt}
\begin{tabular}{lrrrrrrr}
\toprule
Run & Ann.\ return & Ann.\ excess & 95\% CI & IR & Max DD & Years $+$ & DSR$_e$\\
\midrule
weekly, 10\,bps (main) & 26.6\% & $-4.09\%$ & $[-7.73,-0.67]$ & $-0.63$ & 31.2\% & 4/10 & $3\times10^{-5}$\\
weekly, 0\,bps         & 26.8\% & $-3.91\%$ & $[-7.55,-0.51]$ & $-0.60$ & 31.2\% & 4/10 & $4\times10^{-5}$\\
weekly, 25\,bps        & 26.2\% & $-4.36\%$ & $[-8.00,-0.91]$ & $-0.67$ & 31.2\% & 4/10 & $2\times10^{-5}$\\
weekly, relative pressure & 27.6\% & $-3.29\%$ & $[-6.81,-0.03]$ & $-0.51$ & 31.2\% & 4/10 & $1\times10^{-4}$\\
monthly, 10\,bps       & 27.2\% & $-3.62\%$ & $[-7.04,-0.37]$ & $-0.57$ & 31.0\% & 4/10 & $6\times10^{-5}$\\
weekly, top 500        & 25.8\% & $-4.75\%$ & $[-8.48,-1.23]$ & $-0.72$ & 31.2\% & 3/10 & $1\times10^{-5}$\\
\midrule
base, rebalanced weekly & 27.4\% & $-3.47\%$ & $[-6.94,-0.19]$ & $-0.54$ & 31.2\% & 4/10 & --\\
VOO                    & 15.2\% & $-14.88\%$ & $[-24.18,-5.57]$ & $-1.03$ & 34.0\% & 1/10 & --\\
\bottomrule
\end{tabular}
\end{table*}

\begin{figure*}[t]
\centering
\includegraphics[width=0.92\textwidth]{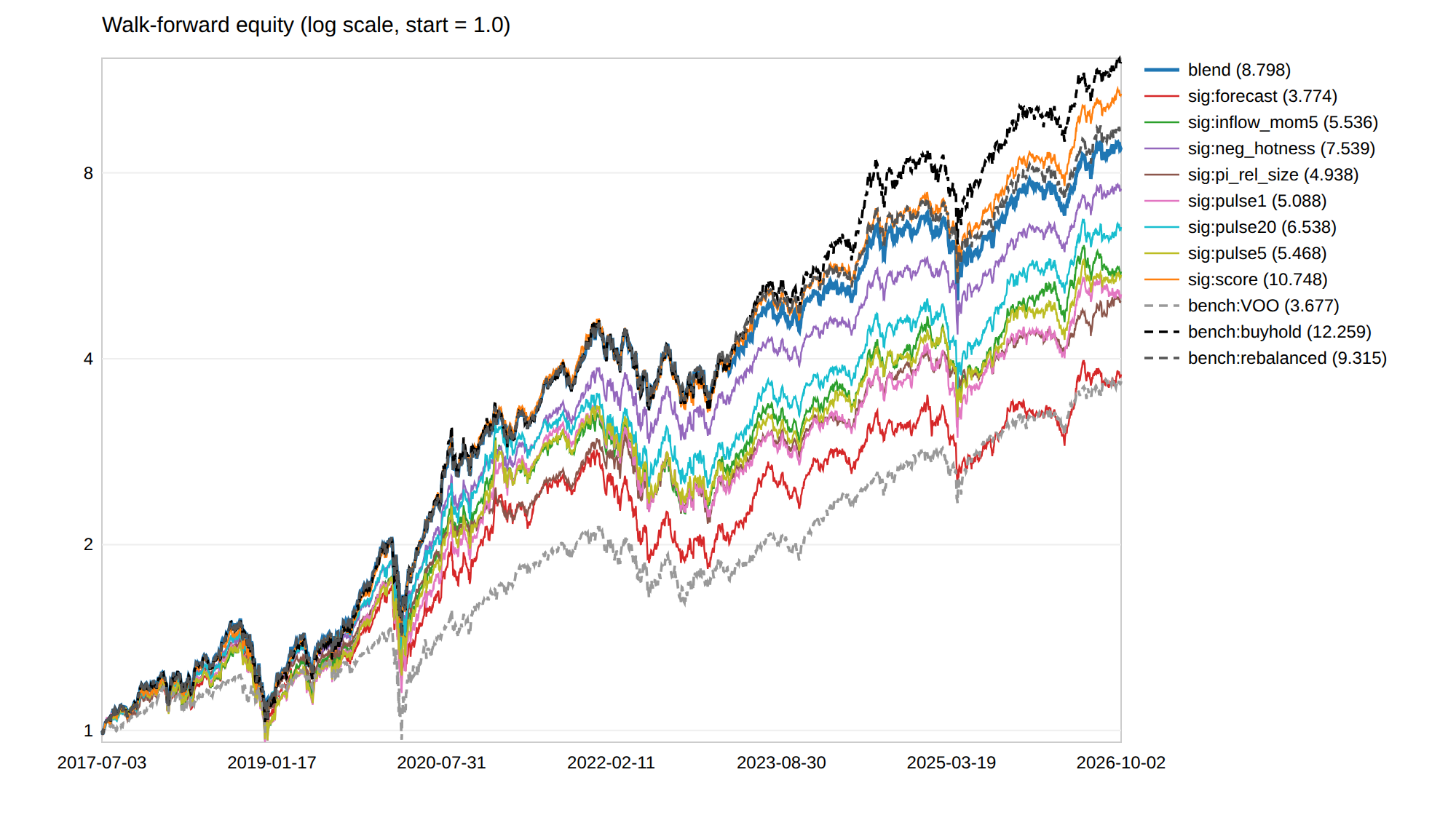}
\caption{Walk-forward equity of the main run (weekly, 10\,bps; log scale, start $=1$). Strategies are solid and benchmarks dashed. Buy-and-hold of the base (black, dashed) ends highest; the score tilt (orange) is the only strategy above the weekly-rebalanced base.}
\label{Fig:wfequity}
\end{figure*}

\paragraph{Decision} On the main run the gate gives: c1 FAIL (excess $-4.09\%$, interval $[-7.73\%,-0.67\%]$); c2 FAIL (4 of 10 years positive); c3 FAIL (deflated IR of the excess $3\times10^{-5}$); c4 pass (drawdown 31.2\% against a 37.2\% limit); c5 FAIL (the large-cap blend fails c1, c2 and c3 as well). \textbf{The answer is ``no''.} As first run, c3 deflated the strategy's total Sharpe ratio and passed (0.989). The final review found that a portfolio that is 80\% the base mix (60\% AAPL) has a high total Sharpe whatever the tilt does, so that version of c3 did not test the excess. The criterion was amended on 2026-10-02 to deflate the information ratio of the excess, and the verdict is ``no'' under either definition. Following the protocol, the work turns to observed flows (M3c) rather than to an optimizer.

\paragraph{Caveats} Two biases favour the strategies. First, the universe is today's ticker list, so stocks delisted since 2016 are missing (survivorship). Second, a holding that stops trading stays frozen at its last price, so a failed pick never shows its final loss; we record this as a known limitation. One bias runs the other way: the base mix is today's portfolio, selected with hindsight and dominated by two of the decade's winners, which biases c1 against any tilt away from it. The negative result is therefore not simply generous to the signals. Two secondaries, reported but not gated and pre-registered for the next experiment, remove the base from the comparison: every strategy against the base rebalanced on the same calendar, and base-free sleeves (the top 10 names at equal weight) against the equal-weight eligible universe. The $t$-statistics of Table~\ref{Table:wfic} ignore cross-sectional dependence and the 32 signal--horizon cells tested, so the IC entries near $|t|=2$ should be read as suggestive at most.

\section{Observed paired flows from SEC Form 13F}
\label{Section:13f}

The flows of Section~\ref{Section:flows} are inferred from bars. SEC Form 13F~\cite{sec13f} gives an \emph{observed}, if coarse, pairing. Every institutional manager with more than \$100M in Section~13(f) securities files its quarter-end long holdings on EDGAR~\cite{edgar}. Between two quarters a manager's cuts are its sources and its adds are its sinks. We build that observed quarterly flow matrix and compare it with the estimate summed over the same quarter's bars.

\paragraph{Method}
\begin{itemize}
\item \textbf{Holdings.} We ingest the SEC's 13F data sets and keep only 13F-HR filings. Per manager and period the latest filing wins: a restatement replaces the holdings, and a new-holdings amendment adds rows. Only share positions count, with options excluded.
\item \textbf{Value repair.} The VALUE field is repaired per manager only, against each CUSIP's consensus price (the median value/shares over at least five holders). A manager is rescaled when it has at least five consensus rows, its median deviation sits at $1000\times$ (or $10^6\times$) within a factor of 3, and at least 60\% of its rows sit at that same power. The consensus is then recomputed without the flagged managers and the test repeated, so a CUSIP held mostly by mis-unit filers cannot drag it. Single rows and share counts are never rewritten: a row far off the consensus is as consistent with a share-count error as with a value error, so it is left for the position guard below. Before the repair the raw totals were \$110--530T per quarter before the 2023 switch from thousands to dollars, and \$32--72T after it. Over all quarters, 1,743 manager-quarters were rescaled down (dollars filed before the switch) and 8,190 up (thousands filed after it).
\item \textbf{Mapping.} CUSIPs map to tickers through OpenFIGI~\cite{openfigi}.
\item \textbf{Observed matrix.} For manager $m$ and stock $i$, $d_{mi}=\Delta\text{shares}_{mi}\,\bar P_{iq}$, where $\bar P_{iq}$ is the quarter's mean lake close, moved onto the filing's share basis by the 13F-median price. A split counts only when the holders' share counts confirm it. Each manager's outflow is paired with its inflow proportionally, $F^{m}_{ij}=\mathrm{out}_i\,\mathrm{in}_j/\sum\mathrm{in}\cdot\min(1,\sum\mathrm{in}/\sum\mathrm{out})$; the remainder is unpaired cash. The observed matrix is $T_q=\sum_m F^m$ on the top 2,000 tickers by 13F value.
\item \textbf{Dropped positions.} A position is dropped when either quarter's value/shares is at least $100\times$ off $\bar P_{iq}$ (the 13F-median price moved through lake returns; $q-1$ at $\bar P_{iq}$ times the split ratio). It is also dropped when a side filed at value 0 holds more than \$1M of shares at that price. Either error would otherwise become a phantom flow.
\item \textbf{Estimate.} The estimate $\hat T_q$ is the exact per-bar flux of Equation~(\ref{eqn:gravity}) summed over the quarter's bars. It is warmed up on at least 60 prior returns and restricted to the same nodes.
\end{itemize}

\paragraph{Data and coverage} The run uses 43 SEC data-set archives (2.65\,GB) covering 42 quarters, 2016Q1--2026Q2:
\begin{itemize}
\item 4,253--8,900 managers and 1.05M--2.40M holdings per quarter, \$20.8T (2016Q1) to \$75.7T (2026Q2) after the unit repair;
\item 23,554 CUSIPs queried on OpenFIGI, covering every CUSIP with a holding of at least \$10M in the first ingest (9 CUSIPs that reach \$10M only after the unit repair were not queried and stay unmapped); in each of the 42 quarters, all top-3,000 CUSIPs by value have such a holding;
\item in the latest quarter, 2026Q2, \textbf{94.2\%} of 13F dollar value maps to a ticker of the current 10,000-ticker universe. The share falls going back (93.1\% in 2024Q4, 87.8\% in 2022Q4, 84.1\% in 2019Q4, 71.9\% in 2016Q1; 87.0\% over all quarters) because the universe holds only today's tickers. One unrepaired \$6.6T row, whose CUSIP has no ticker and so never enters the flows, pulls 2021Q2 down to 73.6\%. Node selection and the 13F-value baseline use values as filed; rows 100x or more off the price, dropped from the flows, change at most 4 of the 2,000 nodes in a quarter pair (checked on 2021Q4/2022Q1 and 2024Q2/Q3).
\end{itemize}
Thirty-nine quarters, 2016Q4--2026Q2, have a previous quarter and enough warm-up. Each has 4,313--9,006 managers, 3.08M--3.71M observed edges and \$0.91T--\$3.37T of paired flow.

\paragraph{Nulls} A lift is the metric minus a null. We use four nulls:
\begin{itemize}
\item the \emph{temporal placebo}, the same configuration's estimate of another quarter ($q-4$, else $q+1$). It cancels the density and size effects of each configuration;
\item the \emph{dense gravity null} $\mathrm{out}\,\mathrm{in}^\top/\text{total}$, built from the estimate's own marginals over every pair with margins;
\item the \emph{support-matched gravity null}: the same product masked to the estimate's own nonzero pairs, with each row rescaled to the estimate's row sum. It has the estimate's density and out-margins and differs only in how each row is split. The edge metrics count missing pairs as 0, so a denser null gains from density alone; this null measures the estimate's pairing beyond its marginals with that effect removed. Per quarter, the share of the $n(n-1)$ pairs that are nonzero is 0.77--0.93 in the observed matrix, 0.20--0.28 in the estimate (0.80M--1.14M edges) and 0.33--0.41 in the dense gravity null;
\item 100 random relabellings of the estimate (\emph{perm}).
\end{itemize}

\paragraph{Results} Table~\ref{Table:13f} reports the base preset ($\lambda=1$, dollar pressure).
\begin{itemize}
\item \textbf{Permutation null.} Every metric clears it by a wide margin.
\item \textbf{Gravity nulls.} The estimate is \emph{below} the dense gravity null on every edge metric: in 39 of 39 quarters on both edge Spearman metrics and 38 of 39 on row cosine for the base preset. Part of that gap is density. On all pairs the lift is $-0.065$ against the dense null but only $-0.004\pm0.001$ against the support-matched null. With density and row sums held fixed, the estimate remains below the support-matched null:
\begin{itemize}
\item observed top-5000 $-0.020\pm0.008$, all pairs $-0.004\pm0.001$, row cosine $-0.047\pm0.019$;
\item negative in 39, 39 and 38 of 39 quarters for the base preset, and in 39 of 39 for the other five configurations (except one quarter of row cosine at $\lambda=1$, $\sqrt{V}$).
\end{itemize}
On $\pi$ the support-matched lift is $-0.000\pm0.002$ (positive in 11 of 39 quarters).
\item \textbf{Placebo.} The lift over the placebo is small but mostly positive. The estimate for the right quarter beats another quarter's estimate by 0.01--0.04 on the edge metrics and 0.045 on $\pi$. The lift is positive in 24--36 of 39 quarters, depending on the metric.
\end{itemize}

\begin{table*}[t]
\centering
\caption{Estimated vs observed (13F) quarterly flows, base preset, mean $\pm$ sd over 39 quarters (2016Q4--2026Q2). Lift = estimate $-$ null. Dense gravity: the estimate's marginal product over all pairs; support-matched: the same product on the estimate's own nonzero pairs, rows rescaled to its row sums. The last column counts the quarters with a positive placebo / support-matched lift.}
\label{Table:13f}
\footnotesize\setlength{\tabcolsep}{4pt}
\begin{tabular}{lcccccc}
\toprule
metric & estimate & lift vs placebo & lift vs dense gravity & lift vs supp.\ gravity & lift vs perm & $>0$ \\
\midrule
edge $\rho$, obs.\ top-5000 & $0.162\pm0.059$ & $+0.036\pm0.032$ & $-0.024\pm0.008$ & $-0.020\pm0.008$ & $+0.163\pm0.058$ & 34 / 0 \\
edge $\rho$, all pairs & $0.357\pm0.019$ & $+0.015\pm0.022$ & $-0.065\pm0.010$ & $-0.004\pm0.001$ & $+0.358\pm0.019$ & 28 / 0 \\
row cosine & $0.182\pm0.043$ & $+0.010\pm0.041$ & $-0.050\pm0.020$ & $-0.047\pm0.019$ & $+0.152\pm0.042$ & 24 / 1 \\
$\pi$ Spearman & $0.613\pm0.031$ & $+0.045\pm0.031$ & $+0.001\pm0.002$ & $-0.000\pm0.002$ & $+0.613\pm0.031$ & 36 / 11 \\
$\pi$ top-50 overlap & $0.432\pm0.064$ & $+0.048\pm0.062$ & $+0.016\pm0.025$ & $+0.011\pm0.023$ & $+0.406\pm0.063$ & 27 / 18 \\
\bottomrule
\end{tabular}
\end{table*}

\paragraph{$\pi$ against size} Observed 13F $\pi$ is mostly size:
\begin{itemize}
\item its Spearman correlation with the quarter's average dollar volume (ADV) is $0.896\pm0.015$, and with 13F value $0.870\pm0.017$;
\item estimated $\pi$ reaches only $0.613$ against observed $\pi$, and $0.687$ against ADV.
\end{itemize}
So plain ADV ranks the observed flow destinations much better than estimated $\pi$ does. Estimated $\pi$ agrees with observed $\pi$ about as well as estimated inflow agrees with observed inflow ($0.621$), and its gravity lift is $0.001$: here $\pi$ is a function of the marginals.

\paragraph{Calibration} The grid crosses $\lambda\in\{0,0.5,1\}$ with dollar and $\sqrt{V}$ pressure.
\begin{itemize}
\item \textbf{Winner.} The best configuration by mean placebo lift on $\pi$ is $\lambda=1$ with $\sqrt{V}$ pressure, at $+0.046\pm0.034$ (range $-0.025$ to $+0.112$). It leads the base preset by 0.002, far inside the quarter-to-quarter spread.
\item \textbf{Other metrics.} The winners differ by metric, all by margins of at most 0.002.
\item \textbf{Dollar vs $\sqrt{V}$.} Dollar pressure has clearly higher raw agreement ($\pi$ $0.613$ vs $0.523$; observed top-5000 edges $0.162$ vs $0.117$).
\item \textbf{$\lambda$.} Within a pressure, it moves $\pi$ by at most 0.043 and the edge Spearman metrics by at most 0.010.
\end{itemize}
The 13F data therefore give no reason to change $\lambda$ or the pressure of the preset.

\paragraph{Split candidates} Price-factor jumps that the holders' share counts do not confirm are treated as no split and listed: 1,249 over 39 quarters, against 606 confirmed splits. The large ones are spin-offs and mergers that the lake's adjustment factors reflect but holders' share counts do not. Examples are APD, MET and HPE (share ratio 1.00), RTX in 2020Q2 (price ratio 0.50, share ratio 1.67) and HLT in 2017Q1. Separately, the position guard dropped 50,915 positions. Their 13F value, \$17.9T, is summed over both quarters' rows as filed, so it double-counts pairs and includes the bogus values. The count spikes to 2,300--3,500 per quarter in 2022Q4--2023Q2 and 2024Q2--Q4. These drops come from a few managers whose filings mix units across rows, which the manager-wide repair deliberately leaves alone; the 2023 switch made this worse. The dollar spikes (\$1.0--2.4T in 2017Q4--2018Q2 and 2021Q4--2022Q2) are a handful of single rows filed $1000\times$ too large.

\paragraph{Limits}
\begin{itemize}
\item \textbf{Close to rank-1 by construction.} Proportional pairing makes $T_q$ a sum of per-manager rank-1 matrices. In practice that sum is close to rank-1: a product of marginals alone matches it better than the estimate does, even on the estimate's own support. Agreement on edges is therefore largely agreement on the marginals, and the comparison cannot reward pairing structure that 13F's own proportional pairing does not contain.
\item \textbf{What 13F leaves out.} The filings are quarterly, long-only, institutional and filed with a 45-day lag. They omit shorts, retail, options and intra-quarter round trips. Fund flows appear as unpaired cash, which in several quarters exceeds the paired flow.
\item \textbf{Prices and survivorship.} Trade prices are unknown, so $d$ uses the quarter's mean close. The universe is today's, which lowers coverage in early years.
\item \textbf{Cleaning heuristics.} The unit repair and the $100\times$ position filter are heuristics against a per-CUSIP consensus.
\end{itemize}

\paragraph{Reading} Measured against observed institutional pairing, the bar-level estimate carries information about \emph{which} stocks send and receive money, and a little about \emph{when}. The placebo lift is small but mostly positive. It carries no information about \emph{who pairs with whom} beyond the marginals. On its own support it is no better than its own marginals, and slightly worse: the support-matched gravity lift is negative on every edge metric, in all 39 quarters except one quarter of row cosine. The size of that deficit is small on all pairs ($-0.004$) and larger where the observed flows are largest ($-0.020$ on the observed top-5000). Even its marginal information is weaker than plain dollar volume.

\section{Learning the connectivity, and the chain at 15 minutes}
\label{Section:learned}

Sections~\ref{Section:wf} and~\ref{Section:13f} locate the weakness in the connectivity of $T$, not in the solver. Three follow-up tests ask whether it can be repaired. Each was registered in advance: the variants, the controls and the pass rule were committed to the repository before the data they judge was used, and every deviation was logged before the run it affected. Table~\ref{Table:learned} summarizes them; all three are negative.

\paragraph{Shared protocol} A small neural model predicts each stock's forward return from its own recent features (returns over several windows, a dollar-volume shock, volatility and pressure) plus a \emph{message} aggregated over a graph. The variants share the encoder, the head and the training, and differ only in that message:
\begin{itemize}
\item \textbf{B0:} no message.
\item \textbf{B0E:} a learned per-stock embedding in place of the message. This identity control absorbs anything a graph could fit about individual stocks without any connectivity.
\item \textbf{learned:} a message over a learned graph.
\item \textbf{Bprior:} a message over \MR's own transition matrix, unchanged.
\end{itemize}
Training is walk-forward with an expanding window, periodic retraining and an embargo. The test statistic is the paired per-date difference in rank IC between two variants, with its $t$-statistic; at 15 minutes the standard error is clustered by trading day. Before any real data, each learned model had to show on synthetic markets with a planted graph that it can find that graph at a realistic signal level.

\paragraph{M4: learning $T$ from scratch, weekly} The learned graph is $S=E_sE_d^\top$, low-rank source and destination embeddings with top-$k$ edges per row, trained end to end on next-week returns. The target was to detect at least 10 of 12 planted synthetic markets (learned beating both B0 and B0E at $t>2$) with every null market clean. After three pre-registered optimization rounds the best configuration detected 9 of 12, and only where the oracle IC of the planted signal was 0.24 or more. That is more than ten times the ICs measured on real weekly data (Table~\ref{Table:wfic}). Per the rule, the model was not run on real data.

\paragraph{M5: learning a correction to $T$, at 15 minutes} Two changes aimed at the power problem. The first is more samples: 15-minute regular-session bars (26 per day) with a 6-bar (1.5-hour) forward return that is masked across the close. The second is a prior: the learned graph starts at \MR's chain and only corrects it, $A_t=\mathrm{softmax}_{\text{row}}\bigl(\log P_t+g(\text{pair})+u_i^\top v_j\bigr)$ over the support of $P_t$, so that at initialization the learned message equals Bprior's. The synthetic intraday markets (300 names, 250 sessions) plant a correction on top of a noisy prior, and learned had to beat both Bprior and B0E at $t>2$ in at least 10 of 12 markets at an oracle IC of 0.05. It detected 2 of 12, and 1 of 12 in a signed variant. The learned correction correlated only 0.10 with the planted one. Where the prior itself was the truth, the signed variant did significantly \emph{worse} than the prior ($t$ down to $-3.6$), which means it overfits. Per the rule, it was not run on real data.

\paragraph{M6: the chain as it is, on real 15-minute data} The last test learns nothing about the graph. The data are the 1,000 most liquid US stocks and ETFs from 2024-10-01 to 2026-10-02: 503 sessions and 12.3 million regular-session bars from the Alpaca SIP feed. At every bar, Bprior's message averages over \MR's kept transition edges computed from bars up to that bar, about 27,000 of them. A new control, \textbf{Bshuf}, uses the same chain with every edge's destination relabeled by one fixed random permutation of the tickers: the same graph shape and weights, attached to the wrong stocks. On a synthetic market whose truth is the chain, Bprior beats Bshuf by $t=6.95$; on a market with no signal, $t=-1.24$. The registered rule required, in both rolling-mode runs (two model seeds), that Bprior beat B0E at $t>2$ with a positive difference in at least two thirds of calendar months, \emph{and} beat Bshuf at $t>2$. It did neither: Bprior $-$ B0E was $-0.0014$ ($t=-0.5$) and $-0.0011$ ($t=-0.5$), positive in 39\% of months, and Bprior $-$ Bshuf was $t=+0.2$ and $+0.9$. A third run, retraining from scratch, agrees ($t=+0.4$ and $-0.1$). Every variant, including B0 with no message, reached the same small IC of about $+0.009$ ($t=2.5$ to $3.8$). That IC comes from the shared stock-level features, not from any graph. The top-minus-bottom decile spread was negative for every variant, about $-0.2\%$ per 1.5 hours after 5\,bps per side.

\begin{table*}[t]
\centering
\caption{Pre-registered tests of the connectivity. Oracle IC is the IC of the planted signal on synthetic markets, the ceiling for any model. Detection means the variant beats every listed control at $t>2$; M6 also requires a positive difference in at least two thirds of months.}
\label{Table:learned}
\small\setlength{\tabcolsep}{4pt}
\resizebox{\textwidth}{!}{%
\begin{tabular}{lllll}
\toprule
Test & Data & Graph in the message & Registered target & Outcome\\
\midrule
M4 & synthetic, weekly, $N=300$ & learned from scratch & $\ge10/12$ detected vs B0, B0E & 9/12, only at oracle IC $\ge0.24$\\
M5 & synthetic, 15\,min, $N=300$ & $T$ plus a learned correction & $\ge10/12$ at oracle IC 0.05 vs Bprior, B0E & 2/12 (signed: 1/12)\\
M6 & real, 15\,min, $N=1000$ & $T$ as is & vs B0E and Bshuf, both seeds & $t=-0.5$ / $+0.2$ and $-0.5$ / $+0.9$\\
\bottomrule
\end{tabular}}
\end{table*}

\paragraph{Reading} The two learned models fail for lack of power. The signal they would have to learn the graph from is far weaker than what these models need, even with 26 times more samples per day and a starting point at \MR's chain. M6 asks the simpler question directly: does the estimated chain carry information about the next 1.5 hours? It does not, beyond what graph-free features already carry, and it does no better than the same chain on the wrong tickers. Together with the 13F comparison, this says that what $T$ knows is in its marginals, at weekly and at intraday horizons alike.

\section{Limitations and future work}
\label{Section:limits}

\paragraph{The flows are inferred} $T$ is an estimate built from simultaneous price and dollar-volume moves (Section~\ref{Section:flows}), not paired transactions. The rank inherits every assumption of Equation~(\ref{eqn:gravity}), and it cannot distinguish a true rotation from two unrelated stocks that moved in opposite directions on the same day.

\paragraph{Teleport floor} Under the \MR\ preset 71\% of active stocks (floor share 0.714) sit at the teleport floor. This comes from the sparsity of the estimate, not from the chain: each source reaches at most 64 of the 256 largest sinks per bar, and two-sided pruning keeps 20 out- and 10 in-edges, so most small stocks never receive kept inflow and hold only teleport mass. The ranking is informative at the top and flat below it. Because floor stocks get the same $\pi$ whatever their size, small floor names read slightly above 0 under the $\log(\pi/s)$ height: a mild relief from teleport mass alone, not a signal. The structure gain of the preset is also low (0.248), so $\pi$ is close to the inflow ranking.

\paragraph{Short-lived communities} The flux communities behind the landscape persist only about as long as the flow window that defines them (Table~\ref{Table:persist}). The territories are a reading aid for the recent flows, not a stable market structure; the exact ranked table is the reference.

\paragraph{Leveraged and inverse products} They are deliberately kept (Section~\ref{Section:etf}), but their flows are partly mechanical: daily rebalancing and retail trading make them flow magnets independent of fundamentals. Their rank should be read as a directional signal about their underlying, not as a quality of the product.

\paragraph{Planned work}
\begin{itemize}
\item \textbf{ETF creation and redemption flows}, which move money into or out of a basket of constituents at once.
\item \textbf{Signed order flow} via the Lee--Ready classification~\cite{leeready} of intraday trades, in place of the sign of the bar return.
\item \textbf{Lagged pairing}, so that money sold on one bar can arrive on a later one.
\item \textbf{Observed edges in $T$}: mixing the 13F pairing (Section~\ref{Section:13f}) into the estimate, at least to set the marginals it already gets right, and testing whether the mixed chain predicts better.
\item \textbf{Live minute streaming}, extending the replay pipeline to append bars as they close.
\end{itemize}

\section{Conclusion}
\label{Section:conclusion}

\MR\ is a Markov-chain transition-probability solver for money: a stock's score is its stationary probability, the share of the market's estimated dollar flow that settles on it in the steady state of a damped chain. PageRank is the best-known instance of the same idea. The model is small enough to check by hand, and the engine ranks 10,000 tickers bit-identically on any thread count. We set out to learn whether it can improve real portfolio moves, fixed the tests in advance, and the answer on public daily bars is no.

\paragraph{It does not beat holding} Over ten years of weekly decisions, after costs and with every configuration registered, following \MR\ lost 3--5\% a year against buy-and-hold of the portfolio, in every run and at every cost level, and failed the pre-registered gate (Section~\ref{Section:wf}). The one persistent effect is a short-term reversal in hotness (IC $+0.006$ for cold names, $t=3.2$, 8 of 10 years), consistent with the price pressure and reversal documented after flow-induced trading~\cite{covalstafford}, but too small to pay for trading. The much larger reversal of the first 60-day evaluation ($t\approx-11$, legacy configuration with relative pressure) did not carry over: with relative pressure the ten-year walk-forward finds $t=1.4$.

\paragraph{The estimated flows are mostly size} Against the observed account-level pairing of SEC Form 13F, 39 quarters and 4,000--9,000 managers each, the estimate gets the senders and receivers roughly right and carries a small timing signal, but nothing about who pairs with whom: on its own support it is slightly worse than a gravity model of its own marginals, in essentially every quarter (Section~\ref{Section:13f}). Plain dollar volume ranks the observed destinations better (0.90) than the estimated $\pi$ does (0.61). The correlation tilt that was meant to supply the pairing does not.

\paragraph{The landscape is honest about its limits} Re-clustered from scratch on every bar, the flux communities persist about as long as the flow window that defines them and are not sectors. The view now shows that churn instead of hiding it; the exact table remains the reference.

\paragraph{Learning the connectivity does not rescue it} Three registered tests asked whether the connectivity can be repaired (Section~\ref{Section:learned}). Learning it from scratch on weekly data, and learning a correction to the chain on 15-minute bars, both failed to recover a planted graph at realistic signal levels on synthetic markets, so neither was run on real data. Tested as it is on two years of real 15-minute data for the 1,000 most liquid names, the chain predicts the next 1.5 hours no better than graph-free controls, and no better than the same chain attached to the wrong stocks.

\paragraph{What would change the answer} The negative results locate the bottleneck in the flows, not the solver: public bars carry size and a little timing, not pairing, and neither learning nor a finer time step recovers the pairing from them. The transition-probability graph remains the core of the method; what must change is the data that set its connectivity. That means data that observe pairing more directly: ETF creation and redemption, signed order flow classified trade by trade rather than by the sign of the bar return, and observed 13F edges mixed into $T$. Each is to be re-tested with the same walk-forward harness, the same 13F nulls and the relabeled-chain control of Section~\ref{Section:learned}.

\section*{Software availability}
\MR\ is a C++20 program, open source under the MIT license at \url{https://github.com/karametebkaan/marketrank}. The model, the evaluation harness and the \FS\ view described here are those of the current implementation. The tests of Section~\ref{Section:learned} are in \texttt{research/graphlearn}, each with its pre-registration and results file.

\bibliography{marketrank}

\begin{thebibliography}{10}
\expandafter\ifx\csname url\endcsname\relax
  \def\url#1{\texttt{#1}}\fi
\expandafter\ifx\csname urlprefix\endcsname\relax\def\urlprefix{URL }\fi
\expandafter\ifx\csname href\endcsname\relax
  \def\href#1#2{#2} \def\path#1{#1}\fi

\bibitem{norris}
J.~R. Norris, Markov Chains, Cambridge Series in Statistical and Probabilistic
  Mathematics, Cambridge University Press, Cambridge, 1997.

\bibitem{kemenysnell}
J.~G. Kemeny, J.~L. Snell, Finite {M}arkov Chains, Van Nostrand, Princeton, NJ,
  1960.

\bibitem{brinpage}
S.~Brin, L.~Page, The anatomy of a large-scale hypertextual {Web} search
  engine, Computer Networks and ISDN Systems 30~(1--7) (1998) 107--117.
\newblock \href {http://dx.doi.org/10.1016/S0169-7552(98)00110-X}
  {\path{doi:10.1016/S0169-7552(98)00110-X}}.

\bibitem{pagerank}
L.~Page, S.~Brin, R.~Motwani, T.~Winograd, The {PageRank} citation ranking:
  Bringing order to the web, Tech. Rep. 1999-66, Stanford InfoLab (1999).

\bibitem{kineticamapmatching}
B.~K. Karamete, L.~Adhami, E.~Glaser, An adaptive {M}arkov chain algorithm
  applied over map-matching of vehicle trip {GPS} data, Geo-spatial Information
  Science 24~(3) (2021) 484--497.
\newblock \href {http://dx.doi.org/10.1080/10095020.2020.1866956}
  {\path{doi:10.1080/10095020.2020.1866956}}.

\bibitem{kineticagraph}
B.~K. Karamete, L.~Adhami, E.~Glaser, \href{https://arxiv.org/abs/2201.02136}{A
  fixed storage distributed graph database hybrid with at-scale {OLAP}
  expression and {I/O} support of a relational {DB}: {Kinetica-Graph}} (2022).
\newblock \href {http://arxiv.org/abs/2201.02136} {\path{arXiv:2201.02136}}.
\newline\urlprefix\url{https://arxiv.org/abs/2201.02136}

\bibitem{kinetica}
Kinetica, Kinetica db. inc document for network graph solvers - v7.0.11,
  \url{https://www.kinetica.com/docs/graph_solver/index.html}, accessed on
  2020-01-06 (2020).

\bibitem{leeready}
C.~M.~C. Lee, M.~J. Ready, Inferring trade direction from intraday data, The
  Journal of Finance 46~(2) (1991) 733--746.
\newblock \href {http://dx.doi.org/10.1111/j.1540-6261.1991.tb02683.x}
  {\path{doi:10.1111/j.1540-6261.1991.tb02683.x}}.

\bibitem{wilson}
A.~G. Wilson, A statistical theory of spatial distribution models,
  Transportation Research 1~(3) (1967) 253--269.
\newblock \href {http://dx.doi.org/10.1016/0041-1647(67)90035-4}
  {\path{doi:10.1016/0041-1647(67)90035-4}}.

\bibitem{louvain}
V.~D. Blondel, J.-L. Guillaume, R.~Lambiotte, E.~Lefebvre, Fast unfolding of
  communities in large networks, Journal of Statistical Mechanics: Theory and
  Experiment 2008~(10) (2008) P10008.
\newblock \href {http://dx.doi.org/10.1088/1742-5468/2008/10/p10008}
  {\path{doi:10.1088/1742-5468/2008/10/p10008}}.

\bibitem{rsb}
Y.~Hu, R.~J. Blake, Numerical experiences with partitioning of unstructured
  meshes, Parallel Computing 20 (1994) 815--829.

\bibitem{fiedler}
M.~Fiedler, Algebraic connectivity of graphs, Czechoslovak Mathematical Journal
  23~(2) (1973) 298--305.

\bibitem{chung}
F.~R.~K. Chung, Spectral Graph Theory, Vol.~92 of CBMS Regional Conference
  Series in Mathematics, American Mathematical Society, 1997.

\bibitem{hilbert}
D.~Hilbert, {\"U}ber die stetige {A}bbildung einer {L}inie auf ein
  {F}l{\"a}chenst{\"u}ck, Mathematische Annalen 38 (1891) 459--460.

\bibitem{gilbert}
J.~{\v{C}}erven{\'y}, gilbert: generalized {H}ilbert (``gilbert'')
  space-filling curve for rectangular domains of arbitrary (non-power of two)
  sizes, \url{https://github.com/jakubcerveny/gilbert}, accessed 2026-10-02.

\bibitem{shepard}
D.~Shepard, A two-dimensional interpolation function for irregularly-spaced
  data, in: Proceedings of the 1968 23rd ACM National Conference, 1968, pp.
  517--524.
\newblock \href {http://dx.doi.org/10.1145/800186.810616}
  {\path{doi:10.1145/800186.810616}}.

\bibitem{cvt}
Q.~Du, V.~Faber, M.~Gunzburger, Centroidal {V}oronoi tessellations:
  applications and algorithms, SIAM Review 41~(4) (1999) 637--676.
\newblock \href {http://dx.doi.org/10.1137/S0036144599352836}
  {\path{doi:10.1137/S0036144599352836}}.

\bibitem{lloyd}
S.~P. Lloyd, Least squares quantization in {PCM}, IEEE Transactions on
  Information Theory 28~(2) (1982) 129--137.
\newblock \href {http://dx.doi.org/10.1109/TIT.1982.1056489}
  {\path{doi:10.1109/TIT.1982.1056489}}.

\bibitem{gouraud}
H.~Gouraud, Continuous shading of curved surfaces, IEEE Transactions on
  Computers C-20~(6) (1971) 623--629.
\newblock \href {http://dx.doi.org/10.1109/T-C.1971.223313}
  {\path{doi:10.1109/T-C.1971.223313}}.

\bibitem{dls}
B.~K. Karamete, R.~Aubry, E.~L. Mestreau, S.~Dey, A novel double link structure
  ({DLS}) with applications to computational engineering and design, AIAA
  Aerospace Sciences Meeting 54 (2016) 1301.
\newblock \href {http://dx.doi.org/10.2514/6.2016-1301}
  {\path{doi:10.2514/6.2016-1301}}.

\bibitem{openmp}
{OpenMP Architecture Review Board}, {OpenMP} application programming interface,
  \url{https://www.openmp.org}, accessed 2026-10-02.

\bibitem{alpaca}
{Alpaca}, Alpaca market data {API} v2: historical stock bars,
  \url{https://docs.alpaca.markets}, accessed 2026-10-02.

\bibitem{parquet}
{Apache Software Foundation}, Apache {Parquet},
  \url{https://parquet.apache.org}, accessed 2026-10-02.

\bibitem{duckdb}
M.~Raasveldt, H.~M{\"u}hleisen, {DuckDB}: an embeddable analytical database,
  in: Proceedings of the 2019 International Conference on Management of Data
  (SIGMOD), 2019, pp. 1981--1984.
\newblock \href {http://dx.doi.org/10.1145/3299869.3320212}
  {\path{doi:10.1145/3299869.3320212}}.

\bibitem{edgar}
{U.S. Securities and Exchange Commission}, {EDGAR}: Electronic data gathering,
  analysis, and retrieval system, \url{https://www.sec.gov/edgar}, accessed
  2026-10-02.

\bibitem{d3}
M.~Bostock, V.~Ogievetsky, J.~Heer, {D}$^3$: Data-driven documents, IEEE
  Transactions on Visualization and Computer Graphics 17~(12) (2011)
  2301--2309.

\bibitem{sec13f}
{U.S. Securities and Exchange Commission}, Form {13F}: reports filed by
  institutional investment managers,
  \url{https://www.sec.gov/divisions/investment/13ffaq.htm}, accessed
  2026-10-02.

\bibitem{openfigi}
{Bloomberg L.P.}, {OpenFIGI} {API} v3 mapping,
  \url{https://www.openfigi.com/api}, accessed 2026-10-02.

\bibitem{covalstafford}
J.~Coval, E.~Stafford, Asset fire sales (and purchases) in equity markets,
  Journal of Financial Economics 86~(2) (2007) 479--512.
\newblock \href {http://dx.doi.org/10.1016/j.jfineco.2006.09.007}
  {\path{doi:10.1016/j.jfineco.2006.09.007}}.

\end{thebibliography}

\end{document}